\documentclass[journal,onecolumn,12pt]{IEEEtran}
\usepackage[top=1in, bottom=1in, left=1in, right=1in]{geometry}
\usepackage{graphicx}
\usepackage{epstopdf}
\usepackage{amsmath,amssymb}
\usepackage{arydshln}
\usepackage{dsfont}
\usepackage{color}
\usepackage{cases}
\usepackage{bm}
\usepackage{amsfonts}

\usepackage[font=footnotesize]{caption}
\usepackage{indentfirst}
\usepackage{cite}

\usepackage{caption}
\usepackage{algorithm}
\usepackage{algpseudocode}
\usepackage{booktabs}
\usepackage{subfigure}
\usepackage{multirow}
\usepackage{amsthm}
\usepackage{setspace}
\usepackage{threeparttable}
\renewcommand{\bm}{\mathbf}
\newcommand{\sd}{\mathbf}
  \newcommand{\sys}{}
\theoremstyle{remark}
\newtheorem{remark}{Remark}
\theoremstyle{}
\newtheorem{theorem}{Theorem}
\theoremstyle{}

\newtheorem{lemma}{Lemma}
\theoremstyle{}
\newtheorem{definition}{Definition}
\theoremstyle{remark}
\newtheorem{example}{Example}

\theoremstyle{definition}

\makeatletter
\newcommand{\tabcaption}{\def\@captype{table}\caption}
\newcommand{\tabincell}[2]{\begin{tabular}{@{}#1@{}}#2\end{tabular}}
\makeatletter
 \allowdisplaybreaks[4]

\ifCLASSINFOpdf
\else
\fi

\definecolor{newcolor}{rgb}{0.5,0,1}

\newcommand{\dt}[1]{{\color{black}#1}}
\newcommand{\yq}[1]{{\color{black}#1}}

\newcommand{\md}[1]{{\color{black}#1}}

\begin{document}

\title{Fundamental Limits of Caching for Demand Privacy against Colluding Users}

\author{
Qifa Yan~\IEEEmembership{Member,~IEEE,} and
Daniela Tuninetti~\IEEEmembership{Fellow,~IEEE.}
\thanks{The Authors are with the Electrical and Computer Engineering Department of the University of Illinois Chicago, Chicago, IL 60607, USA (e-mail: qifa@uic.edu, danielat@uic.edu). This work was supported in part by NSF Award 1910309.}
}
\maketitle


\begin{abstract}
This work investigates the problem of \emph{demand privacy against colluding users} for shared-link coded caching systems, where no subset of users can learn any information about the demands of the remaining users. The notion of privacy used here is stronger than similar notions adopted in past work and is motivated by the practical need to insure privacy regardless of the file distribution. 
\yq{
Two scenarios are considered: \emph{Single File Retrieval (\dt{SFR})} and \emph{Linear Function Retrieval (LFR)}, where in the latter case each user demands an arbitrary linear combination of the files at the server. 
}
The main contributions of this paper are a novel achievable scheme for LFR, referred as \emph{privacy key scheme}, 
and a new information theoretic converse bound for SFR. 
\dt{Clearly, being SFR a special case of LFR,  an achievable scheme for LFR works for SFR as well, and  a converse for SFR is a valid converse for LFR as well.}
By comparing the performance of the achievable scheme with the converse bound derived in this paper (for the small cache size regime) and existing converse bounds without privacy constraints \dt{(in the remaining memory regime)}, the communication load of the privacy key scheme turns out to be optimal to within a constant multiplicative gap in all parameter regimes. 
Numerical results show that the new privacy key scheme outperforms in some regime known schemes based on the idea of virtual users, which also satisfy the stronger notion of user privacy against colluding users adopted here. Moreover, the privacy key scheme enjoys much lower subpacketization than known schemes based on virtual users.
\end{abstract}


\section{Introduction}

Coded caching is a promising technique to reduce network congestion during peak times. Consider a shared-link network consisting of a server having access to a library of $N$ files and being connected to $K$ users, each equipped with a local cache memory of size $M$ files. The network operates in two phases. The \emph{placement phase} happens when the network is not congested, during which the server pushes some content into each user's local cache without knowing their future demands. \dt{Placement is said to be uncoded if all cache contents simply contain copies of some bits of the files at the server.} The \emph{delivery phase} happens at peak times, during which each user demands one file\footnote{Single File Retrieval (SFR) is the classical coded caching problem formulation~\cite{Maddah-Ali2014fundamental}, recently extended to Linear Function Retrieval (LFR) in~\cite{Kai2020LinearFunction}.}
 from the server and the server responds by sending a signal to satisfy the users' demands. In coded caching, the \emph{worst-case communication load} (or just \emph{load} for short in the following) is reduced by creating multicast opportunities in the delivery phase by cleverly pushing content in the caches in the placement phase. 

The \md{technique  of coded caching} was introduced by Maddah-Ali and Niesen (MAN) in~\cite{Maddah-Ali2014fundamental}. The MAN scheme was proved to achieve the optimal  load among all uncoded placement schemes when $N\geq K$ in~\cite{Kai2020Index}. By removing redundant transmissions in the MAN scheme, the optimal load-memory tradeoff among all uncoded placement schemes was characterized for $N < K$ in~\cite{Yu2019ExactTradeoff}. The schemes in~\cite{Maddah-Ali2014fundamental,Yu2019ExactTradeoff} allow each user to decode an arbitrary file from the library and will be referred to as Single File Retrieval (SFR) schemes in the following. 
The load-memory tradeoff of SFR in~\cite{Yu2019ExactTradeoff} was showed to be achievable also for Linear Function Retrieval (LFR) with uncoded placement in~\cite{Kai2020LinearFunction}, where each user is allowed to demand an arbitrary linear combination of the files at the server.  Improved achievable loads by using coded placement were obtained in~\cite{code:prefetch01,code:prefetch02,code:prefetch03} and the original cut-set converse bound in~\cite{Maddah-Ali2014fundamental} was improved upon in~\cite{Ramamoorthy2017bound,ImproveBound}; we know however that uncoded placement for SFR is no more than a factor of $2$ away from optimal~\cite{QYu2018Factor2}.

\md{
Protecting demand privacy is an important aspect in practical systems.  Private Information Retrieval (PIR) was first proposed by Chor et al in~\cite{BhorPIR1,BhorPIR2} to address this issue. In the basic PIR setup, there are multiple non-colluding servers, each storing the same library of files, and a single user;
the user aims to retrieve a single file from the servers while keeping the index of the demanded file private from any individual server. The capacity of basic PIR was characterized in~\cite{HuaPIR}. Since~\cite{HuaPIR} there has been a great interest in studying variations of the basic PIR model, such as PIR against colluding servers~\cite{SunT,YaoT}, PIR against coded databases~\cite{PIRcoded_database1,PIRcoded_database2,PIRcoded_database3} or PIR when the user has a local memory or side information~\cite{CachePIR,PIRsingle}.  
Privacy in PIR protects the user's demand against the servers. In single-server shared-link networks, there are multiple users with a local cache but only a single server, thus demand privacy is intended to protect the identity of the demanded file by a user against the other users, e.g.,  index coding~\cite{Song}.
}

In coded caching, decoding the MAN multicast signals requires global knowledge of the demands of the users, thus infringing the privacy of the users. Moreover, users may learn the content of files other than the requested one, thus threatening security.
Information theoretic \emph{secure} coded caching was considered in~\cite{secure,Privacy} for the shared link system. In~\cite{secure}, secure delivery was investigated,  where a wiretapper who observes the transmitted signal can not learn any information about the files; in~\cite{Privacy}, secure caching was investigated, where each user can not obtain any information on non-demanded files. \md{Secure delivery and caching are simultaneously considered in device-to-device~\cite{Zewail:D2D} and combination networks~\cite{Zewail:Combination}.}
Information theoretic \emph{demand-private} coded caching was formalized in~\cite{Kai2019Private}, where the aim is to guarantee that each user does not learn any information about the indices of the files demanded by the other users.
Relevant for this work is a way to ensure privacy informally referred to as scheme with \emph{virtual users}, an idea that first appeared in~\cite{EliaPrivay} and was later analyzed in~\cite{Kai2019Private,Kamath2019}.
The idea is that a private scheme for a system with $K$ users and $N$ files can be constructed from known non-private schemes (such as~\cite{Maddah-Ali2014fundamental,Yu2019ExactTradeoff}) for $NK$ users and $N$ files, that is, by introducing $K(N-1)$ virtual users.
In the placement phase of a virtual user scheme, the users choose their cache contents from the $NK$ caches of the non-private scheme without replacement, privately and randomly; in the delivery phase, the demands of the $K$ users are extended to demands for $NK$ users (including $K$ real users and $N(K-1)$ virtual users) such that each file is demanded exactly $K$ times. The server sends multicast signals to satisfy the extended demands of $NK$ users according to the non-private coded caching scheme. Privacy of the real users is guaranteed since each real user can not distinguish the demands of real and virtual users. The general idea of transforming a non-private coded caching scheme for $NK$ users to a private one for $K$ users was further studied in~\cite{Aravind2019,Sneha2019}, and later extended to device-to-device network in~\cite{KaiD2DPrivacy}, where a trusted server having no access to the file library coordinates the transmission among the users.

In this paper, we investigate the problem of \emph{Demand Privacy against Colluding Users (DPCU)} for both SFR and LFR.
 In DPCU, privacy is guaranteed in the  sense that any subset of colluding users, who may share their cache contents, can not learn any information about the indices of the files demanded by the remaining users. 
It was noted already in~\cite{KaiD2DPrivacy} that the virtual user scheme in~\cite{Kai2019Private,Kamath2019}, which were not designed to fight colluding users, are indeed private against colluding users as well. In this paper we further strengthen the privacy notion of~\cite{KaiD2DPrivacy} by imposing that a feasible scheme should guarantee privacy for \emph{any} file realization. This notion of privacy is motivated by the fact that, in practical systems, the distribution of the files is usually hard to characterize, or even known; in other words, files may not be identically and uniformly distributed, as many theoretical models assume. 
\md{
Remarkably, it was showed in~\cite{Sneha2019} that the load $R=\min\{N,K\}$ is achievable for SFR under the previously adopted privacy condition even without local cache availability, that is, when $M=0$. We will see that with our more stringent definition of privacy the optimal load at $M=0$ for SFR (and thus also for LFR) must satisfy $R\geq N$. With our new converse we thus able to show that virtual user schemes, which satisfy our new privacy definition, are optimal at $M=0$ under the new privacy definition. One practical challenge in designing coded caching schemes is the high subpacketization, defined as the minimum file length needed to realize the scheme. It is well known that, with fixed number of files and cache size at each user, the subpacketization of MAN-type schemes increase exponentially with the number of users $K$~\cite{FiniteLength2016,Yan2017PDA}. This problem is more prominent in virtual user schemes, since they are based on non-private schemes for $NK$ users.} In this work we aim to deign schemes that guarantee DPCU and at the same time have a lower subpacketization than virtual user schemes.

\subsection{Paper Contributions}
The main contributions of this paper are as follows.
\begin{itemize}

\item 
We propose an achievable DPCU scheme, referred to as \emph{privacy key}, for both SFR and LFR. 
Our privacy key scheme leverages the non-private LFR in~\cite{Kai2020LinearFunction}, even for SFR. 
The privacy key starts with the same file split and placement strategy as in MAN~\cite{Maddah-Ali2014fundamental,Kai2020LinearFunction}. In addition, each user privately caches a key that is formed as a random linear combination of the uncached subfiles under the MAN strategy. In the delivery phase, the server broadcasts multicast signals so that each user can decode a linear combination of the subfiles, which can be thought of as containing a desired subfile protected by the local private key.  
\yq{
The difference between the SFR scheme and the LFR scheme is that the privacy key for SFR \dt{can be restricted to lie in a linear subspace (i.e., the random coefficients satisfy a linear relationship),} which results in a load reduction in some regimes compared to the privacy key scheme for LFR.
}

\item 
We derive an information theoretic converse bound for SFR demands under the new privacy definition, which outperforms known bounds for the small memory regime when $N>K$ \dt{and is not restricted to uncoded placement. The new converse for SFR is trivially also a converse for LFR.} The converse is inspired by the approach in~\cite{Sneha2019}, which characterized the exact load-memory tradeoff for the case $N=K=2$. 
\md{
In particular, we derive a lower bound on the sum of conditional entropies, where each entropy term includes a subset of  sent signals and cache contents and is conditioned on some demands, which in turn provides a bound on a weighted sum of the load and the memory size.
We combine the different entropy terms, i.e., different subsets of sent signals and cache contents, by using the submodularity of entropy so as the largest number of files is determined by the combined set of sent signals and cache contents.
}


\item 
By using our new converse bound with privacy combined with known converse bounds without privacy constraints~\cite{QYu2018Factor2}, we show that our privacy key 
schemes are optimal to within a constant multiplicative gap in all parameter regimes. Numerical results indicate that the privacy key scheme  outperforms the virtual user  scheme in~\cite{Kamath2019} for SFR demands  when the library size $N$ is larger than $2K+1$ and the memory size $M$ is smaller than $N-1-\frac{1}{K}$. Thus, even in the stronger sense of privacy used here, the virtual users scheme can be improved upon when $N$ is much larger than $K$.

It is also worth pointing out the superiority of our privacy key schemes compared to virtual user schemes in terms of subpacketization. In contrast to virtual user scheme whose complexity is exponential in $KN$, the subpacketization of our privacy key schemes does not increase compared to the non-private MAN scheme since no virtual users are introduced. In other words, we show that privacy needs not come at the expense of subpacketization.

\end{itemize}

\subsection{Extensions}
\yq{
Since the submission of this work, progress has been made to extend the setup studied here to more general scenarios. 
In our recent paper~\cite{SP:LFR}, we showed that privacy keys can be used together with security keys~\cite{secure} in a structured superposition way to guarantee both content security and demand privacy against colluding users simultaneously. Surprisingly, we showed that content security comes without any penalty on either the load-memory tradeoff or the subpacketization level compared to schemes that only guarantee content security with security keys~\cite{secure}.

Ongoing work includes an extension for the privacy \& security key scheme in~\cite{SP:LFR} to systems with 
multiple colluding servers, where the files are stored accross the servers in coded form. 
Some recent progress for multiple servers was recently presented at~\cite{XZhangMultiUser1,XZhangMultiUser2} for SFR demands.
More generally, we are interested in characterizing the optimal performance of caching systems with LFR demands where user-privacy (as studied in this paper) and server-privacy (as in PIR) are simultaneously present and where any subsets of users or servers can collude. 
Developing such a general framework is part of ongoing work.
}

\subsection{Paper Organization}
The paper is organized as follows.
Section~\ref{sec:model} introduces the problem formulation and
Section~\ref{sec:LFR:CPA} presents the privacy key schemes.
Section~\ref{sec:converse} contains the derivation of our new converse bound and shows that privacy key schemes are order optimal.
Section~\ref{sec:numerical} presents some numerical results. Finally,
Section~\ref{sec:conclusion} concludes the paper. Some proofs can be found in Appendix.

\subsection{Notation}
In this paper, we use  $\mathbb{R}_+$ to denote the set of  non-negative real numbers, and $\mathbb{F}_q$ to denote the finite field of size $q$ where $q$ is a prime power. For a positive integer $n$, $\mathbb{F}_q^n$ is the $n$ dimensional vector space over the field $\mathbb{F}_q$, and $[n]$ is the set of the first $n$ positive integers $\{1,\ldots,n\}$. For a sequence of variables 
$Z_1,\ldots,Z_n$ and an index set $\mathcal{S}\subseteq[n]$, we use the notation $Z_{\mathcal{S}} := \{Z_{i}:i\in\mathcal{S}\}$ and $Z_{[0]} := \emptyset$. 
For integers $m,n$, we use ${n\choose m}$ to denote the binomial coefficient   $\frac{n!}{m!(n-m)!}$, and adopt the convention ${n\choose m}=0$ if $m>n$. The notation $\oplus$ is used to denote the Exclusive OR (XOR) operation. Thoughout the paper, we will not distingush the multiplication and addition operations on the finite field from the real field if the context is clear.


\section{System Model}
\label{sec:model}

Let $N,K$ be positive integers.
An $(N,K)$ system with Demand Privacy against Colluding Users (DPCU), consists of a server with $N$ files, denoted as $W_1,\ldots,W_N,$ and $K$ users, denoted as $1,\ldots,K$, where the server is connected to the users via an error-free shared link. The $N$ files are independently and uniformly distributed over $\mathbb{F}_q^B$ for some prime power $q$ and some integer $B$\footnote{Our achievable scheme works for arbitrary distribution of $W_{[N]}$, but the converse relies on the assumption of independently and uniformly distributed files.}.
Each user $k\in[K]$ has a memory to cache some of the files.
Since there is no problem of protecting  privacy for either $N=1$ or $K=1$, we assume in the following that $N\geq 2$ and $K\geq 2$.  
The system operates in two phases as follows.

\textbf{Placement Phase:}  
The server privately generates a random variable $\sys{P}$ from some probability space $\mathcal{P}$ and fills the cache of each user $k\in[K]$ by using the \emph{cache function}
\begin{IEEEeqnarray}{c}
\md{ \varphi_k:\mathcal{P}\times\mathbb{F}_q^{NB}\mapsto\mathbb{F}_q^{\lfloor MB \rfloor},}
\end{IEEEeqnarray}
for some $M\in[0,N]$, where $M$ is referred to as the cache size or memory size.
The cached content of user $k\in[K]$ is denoted by
 \begin{IEEEeqnarray}{c}
 \sys{Z}_k=\varphi_k(\sys{P},\sys{W}_{[N]}),\quad \forall\, k\in[K].\label{eqn:Zk}
  \end{IEEEeqnarray}
\md{
We assume that the functions $\varphi_1,\ldots,\varphi_K$ are known to the server and all users, but the randomness $P$ is not available at the users except through the cache function in~\eqref{eqn:Zk}, that is, if any randomness is needed by user $k\in[K]$, it must be stored in or computed from $Z_k$.}


\textbf{Delivery Phase:} 
User $k\in[K]$ demands $\sd{d}_k=(d_{k,1},\ldots,d_{k,N})^\mathsf{T}$ from the server, where $\sd{d}_1,\ldots,\sd{d}_K$ are independently distributed over some subset  $\mathcal{D}\subseteq\mathbb{F}_q^N$, which means that user $k$ is interested in retrieving the linear combination of the files
\begin{IEEEeqnarray}{c}
W_{\sd{d}_k} := \sum_{n\in[N]} d_{k,n}\cdot W_n.\label{eqn:line:W}
\end{IEEEeqnarray}
The files $W_{[N]}$, the randomness $P$ and the demands $\sd{d}_{[K]}$ are independent, that is,
\begin{IEEEeqnarray}{c}
H(\sd{d}_{[K]},P,W_{[N]}) = \sum_{k\in[K]}H(\sd{d}_k)+H(P)+\sum_{n\in[N]}H(W_n).
\label{eqn:indep.cond.}
\end{IEEEeqnarray}
In this paper, logarithms are in base $q$.

Given the demands, the server creates the signal $\sys{X}$ by using the \emph{encoding function}
\begin{IEEEeqnarray}{c}
\md{\phi:  \mathcal{P}\times\mathcal{D}^K\times \mathbb{F}_q^{NB}\mapsto\mathbb{F}_q^{\lceil RB \rceil},}\label{eqn:encode}
\end{IEEEeqnarray}
for some $R\geq 0$, where $R$ is referred to as load of the system.
The server transmits to the users via the shared link the signal
\begin{IEEEeqnarray}{rCl}
\sys{X}&=&\phi(\sys{P},\sd{d}_{[K]},\sys{W}_{[N]}).
\end{IEEEeqnarray}


Each user $k\in[K]$ must decode its demanded linear combination $W_{\sd{d}_k}$ in~\eqref{eqn:line:W} by using $(X,Z_k)$, and privacy must be guaranteed against  colluding users, that is, a working  scheme requires
\begin{IEEEeqnarray}{CrCl}
\textnormal{[Correctness]}\quad&H(W_{\sd{d}_k}\,|\,\sys{X},\sd{d}_k,\sys{Z}_k)&=&0,\quad\forall\,k\in[K],\label{eqn:correctness}\\
 \textnormal{[Privacy]}\quad&I(\sd{d}_{{[K]\backslash\mathcal{S}}};\sys{X},\sd{d}_\mathcal{S},\sys{Z}_\mathcal{S}\,|\,\sys{W}_{[N]})&=&0,\quad\forall\,\mathcal{S}\subseteq[K],\mathcal{S}\neq \emptyset.\label{eqn:privacy}
\end{IEEEeqnarray}

\yq{
\begin{definition}\label{def:best}
A memory-load pair $(M,R)\in\mathbb{R}_+^2$ is said to be \dt{$\mathcal{D}$-achievable} if there exist cache and encoding functions such the correctness and privacy conditions are satisfied with memory size $M$ and load $R$ for all demands in $\mathcal{D}$.
The optimal load-memory tradeoff for demands in $\mathcal{D}$ is defined as
\begin{IEEEeqnarray}{c}
R_\mathcal{D}^*(M)=\liminf_{B\rightarrow\infty}\{R:(M,R)~\textnormal{is \dt{$\mathcal{D}$-achievable for some $B$}}\}. 
\end{IEEEeqnarray}
\end{definition}

In this paper, we consider the following two cases for the dead set $\mathcal{D}$:
\begin{enumerate}

\item \emph{Single File Retrieval (SFR):} 
Let $\bm{e}_n, n\in[N],$ denote the $n$-th standard unit vector in $\mathbb{F}_q^N$, i.e.,  the vector with the $n$-th entry being one and all other entries being zeros.
When $\mathcal{D}=\{\bm{e}_1,\ldots,\bm{e}_N\}$, each user wants to decode a single file, since  $W_{\bm{e}_n}=W_n$ for all $n\in[N]$.  
The optimal load $R_{\{\bm{e}_1,\ldots,\bm{e}_N\}}^*$ will be indicated as $R_{\rm{F}}^*$ for short in the following.

\item \emph{Linear Function Retrieval (LFR):} 
When $\mathcal{D}=\mathbb{F}_q^{N}$, the system allows the users to decode any linear combination of the files. 
The optimal load $R_{\mathbb{F}_q^{N}}^*$ will be indicated as $R_{\rm{L}}^*$ for short in the following.

\end{enumerate}
Trivially, since the SFR demands are a subset of the LFR demands (i.e., $\{\sd{e}_1,\ldots \sd{e}_N\}\subseteq\mathbb{F}_q^N$), 
let $R_{\rm{F, converse}}(M)$ be a lower bound for $R_{\rm{F}}^*(M)$, and $R_{\rm{L, achievable}}(M)$ be an achievable load-memory tradeoff  for LFR demands,
it must hold 
\begin{IEEEeqnarray}{c}
\md{R_{\rm{F, converse}}(M) \leq R_{\rm{F}}^*(M) \leq R_{\rm{L}}^*(M) \leq R_{\rm{L, achievable}}(M), \quad\forall M\in[0,N].}
\label{eq:TrivialRelationships}
\end{IEEEeqnarray} 
The main result of this paper is to derive novel $R_{\rm{F, converse}}(M)$ and $R_{\rm{L, achievable}}(M)$ such that $\sup_{M,K,N} \frac{R_{\rm{L, achievable}}(M)}{R_{\rm{F, converse}}(M)}$ is upper bounded by a constant, that is, we design a DPCU scheme that is optimal to within a constant gap for cache aided linear function retrieval in all memory regimes, for any number of users and files.
}


\begin{remark}[\dt{Transmit signal and demands are independent}]\label{remark:privacy}
The intuition behind the privacy guarantee in~\eqref{eqn:privacy} is that, for any file realization $\sys{W}_{[N]}=w_{[N]}$ and for any non-empty set of users $\mathcal{S}\subseteq[K]$, the colluding users in $\mathcal{S}$ can not learn any information on the demands of the other users. Thus, \emph{the privacy is guaranteed irrespective of file realizations and the subset of users participating the collusion.}
%
In Appendix~\ref{sec:app:IDX}, we show that, the condition~\eqref{eqn:privacy} implies  $I(\sd{d}_{[K]}; X\,|\,W_{[N]})=0$, i.e., the equality in~\eqref{eqn:privacy} also holds for $\mathcal{S}=\emptyset$. \emph{This indicates that the signal $X$ is independent of the demands $\sd{d}_{[K]}$, irrespective of the file realization.}
By the fact that the demands $\sd{d}_{[K]}$ are independent of the files $W_{[N]}$ (see~\eqref{eqn:indep.cond.}), the privacy condition in~\eqref{eqn:privacy} is equivalent to
\begin{IEEEeqnarray}{c}
I(\sd{d}_{[K]\backslash\mathcal{S}}\,;\,\sys{X},\sd{d}_\mathcal{S},\sys{Z}_\mathcal{S},\sys{W}_{[N]})=0,\quad\forall\,\mathcal{S}\subseteq[K].\label{eqn:privacyII}
\end{IEEEeqnarray}
\end{remark}

\begin{remark}[\dt{Privacy notions}]
The privacy notions used in~\cite{Kai2019Private,Kamath2019,Sneha2019} and~\cite{Aravind2019} for SFR demands differ from the one used here, in particular 
\begin{IEEEeqnarray}{ccc}
\textnormal{[Privacy in~\cite{Kai2019Private,Kamath2019,Sneha2019}]}&\quad& I(\sd{d}_{{[K]\backslash\{k\}}};\sys{X},\sd{d}_k,\sys{Z}_k)=0,\quad \forall\, k\in[K],\label{other:privacy:1}\\
\textnormal{[Privacy in~\cite{Aravind2019}]}&\quad &I(\sd{d}_j; \sys{X},\sd{d}_k,\sys{Z}_k)=0,\quad \forall\, (j,k)\in[K]^2:j\neq k.\label{other:privacy:2}
\end{IEEEeqnarray}
The above two definitions are obviously implied by our equivalent privacy condition in~\eqref{eqn:privacyII}, \dt{that is, our privacy definition in~\eqref{eqn:privacy} is stronger than any other definition used in past work.}

The definitions of privacy in~\eqref{other:privacy:1},~\eqref{other:privacy:2}  involve one user at a time, that is, users are not assumed to be able to collude.  Demand privacy against colluding users was first introduced in the device-to-device setup~\cite{KaiD2DPrivacy}, where the privacy condition there was defined without conditioning on the library files $W_{[N]}$. The following Example~\ref{exam:1} from~\cite{Sneha2019} illustrates that the condition in~\eqref{other:privacy:1} 
can not guarantee privacy for arbitrary distribution of $\sys{W}_{[N]}$.
\end{remark}

\begin{example}[An SFR scheme from~\cite{Sneha2019} for $N>K$]\label{exam:1}
Consider an $(N,K)$ system where $N>K$ and the files are uniformly and independently distributed over $\mathbb{F}_2^B$. Let $M\in[0,N]$.

In the placement phase, the server generates
\begin{IEEEeqnarray}{c}
P=(T_1,\ldots,T_K,S_1,\ldots,S_K,V_1,\ldots,V_K),
\end{IEEEeqnarray}
where $(T_1,\ldots,T_K)$ is a random permutation of $[K]$, which is uniformly drawn from all permutations of the set $[K]$; the $K$ identically and independently  distributed (i.i.d.) random  variables $S_1,\ldots,S_K$  are uniformly drawn from $[K]$; finally, $V_1,\ldots,V_K$ are i.i.d.  random variables uniformly drawn from $\mathbb{F}_2^{(1-M/N)B}$. The three parts $T_{[K]},S_{[K]}$ and $V_{[K]}$ are independently generated.
Each file is split into two parts as $\sys{W}_n=(\sys{W}_n^{\rm{(c)}},\sys{W}_n^{\rm{(u)}}), n\in[N]$, where $\sys{W}_n^{\rm{(c)}}$ is of size $\frac{M}{N}B$, and $\sys{W}_n^{\rm{(u)}}$ is of size $(1-\frac{M}{N})B$. Each user $k\in[K]$ caches $Z_k=\big(S_k,W_{[N]}^{\rm{(c)}}\big)$, where we refer to $S_k$ as the key of user $k\in[K]$.

In the delivery phase, for demands $\sd{d}_1,\ldots,\sd{d}_K\in \{\bm{e}_1,\ldots,\bm{e}_N\}$, the server first generates a sequence of numbers $J_1,\ldots,J_K$ inductively as follows
\begin{IEEEeqnarray}{c}
\sys{J}_i=\left\{\begin{array}{ll}
                   \sys{J}_j, &\text{if}~\sd{d}_i=\sd{d}_j~\text{for some } j<i \\
                   T_i, &\text{if}~\sd{d}_i\neq\sd{d}_j,\,\forall\,j<i
                 \end{array},
\right.
\end{IEEEeqnarray}
and then sends a signal $\sys{X}=(\sys{Q}_{[K]},\sys{Y}_{[K]})$, where $Q_{[K]}$ \md{and $Y_{[K]}$} are recursively generated as
 \begin{IEEEeqnarray}{rCl}
Q_j&=&(J_j+S_j)_{(K)},\quad \forall\, j\in[K],\\
 \sys{Y}_j&=&\left\{\begin{array}{ll}
                     W_{\sd{d}_i}^{\rm{(u)}},  &\md{\text{if}~j=\sys{T}_i~\text{for some}~i\in[K]}  \\
                     V_j, &\text{otherwise}
                    \end{array}
 \right.,\forall\, j\in[K],
\end{IEEEeqnarray}
where $(\cdot)_{(K)}$ is the modulo operation defined as $(m K+j)_{(K)}=j$ for $j=1,\ldots,K$ and any integer $m$.
The ``side information'' $\sys{J}_k$  records the position of the packet $W_{\sd{d}_k}^{\rm{(u)}}$ in $(Y_1,\ldots,Y_K)$. Since user $k$ has the key $\sys{S}_k$, it can find the position of packet $W_{\sd{d}_k}^{\rm{(u)}}$ from $\sys{Q}_k$, and hence it can decode its missing packet $W_{\sd{d}_k}^{\rm{(u)}}$. Moreover, it was proved in~\cite{Sneha2019} that $I(\sd{d}_{{[K]\backslash\{k\}}};\sys{X},\sd{d}_k,\sys{Z}_k)=0$, where the proof relies on the fact that $(Y_1,\ldots,Y_K,W_1^{\rm{(c)}},\ldots,W_N^{\rm{(c)}})$ is uniformly distributed over $\mathbb{F}_q^{MB+K(1-M/N)B}$, and is independent of $(\sd{d}_{[K]}, Q_{[K]},S_k,J_k)$ for each $k\in[K]$.

Intuitively, the privacy guarantee relies on the fact that the user $k$ can not distinguish if the signal $\sys{Y}_j$ is a random generated vector $V_j$  or some partial file in $W_{[N]\backslash\{n\}}^{\rm{(u)}}$ where $\sd{d}_k=\bm e_n$, since both of them are independently and uniformly distributed over $\mathbb{F}_2^B$, and independent of the cached packets $W_{[N]}^{\rm{(c)}}$. In many practical applications, true file distributions maybe unavailable,  or even they are available, the $N$ files may have different distributions or not independent of the cached packets $W_{[N]}^{\rm{(c)}}$. 
\md{For example, consider the case that bits of file $W_n$ are i.i.d. random variables with Bernoulli distribution with parameter $p_n\in[0,1]$, where  the parameters $p_1,\ldots,p_N$ are distinct. In this case, for each $n\in[N]$, any user can guess each component $Y_j$ by conducting a hypothesis test to determine whether it is the packet $W_n^{(\rm u)}$ or not.} Obviously, the scheme does not satisfy our privacy condition in~\eqref{eqn:privacy}.
\end{example}

\section{Privacy Key Schemes}
\label{sec:LFR:CPA}

In this section we propose an achievable scheme for SFR first, referred to as \emph{privacy key}. Then we modify the scheme to allow for more general LFR demands.
The results are stated in Theorem~\ref{thm:1} and~\ref{thm:2} in Section~\ref{sec:thms}. Before proving those theorems, we provide an illustrative example to highlight the key ingredients of the privacy key schemes in Section~\ref{sec:illustrative example}. Finally, we describe and analyze the general privacy key for SFR and LFR demands in Sections~\ref{sec:scheme} and~\ref{subsec:LFR}, respectively. Some numerical results are presented in Section~\ref{sec:numerical}.

\subsection{Main Results}\label{sec:thms}
\begin{theorem}[SFR]\label{thm:1}
For an $(N,K)$ DPCU system with SFR demands, the lower convex envelope of the memory-load pairs in $\{(0,N)\}\cup\{(M_t,R_t):t\in[0:K]\}$ is achievable, where
\begin{IEEEeqnarray}{c}
(M_t,R_t) := \bigg(1+\frac{t(N-1)}{K},\frac{{K\choose t+1}-{K-\min\{N-1,K\}\choose t+1}}{{K\choose t}}\bigg),\quad t\in[0:K].\label{eqn:achievable}
\end{IEEEeqnarray}
Moreover, the point $(0,N)$ can be achieved with subpacketization $1$, and the point $(M_t,R_t)$
can be achieved with subpacketization $F_t := {K\choose t}, \ t\in[0:K]$.
\end{theorem}

\yq{
\begin{theorem}[LFR]\label{thm:2}
For an $(N,K)$ DPCU system with LFR demands, the lower convex envelope of the memory-load pairs  in $\{(0,N)\}\cup\{(M_t,R_t'):t\in[0:K]\}$  is achievable, where
\begin{IEEEeqnarray}{c}
(M_t,R_t') := \bigg(1+\frac{t(N-1)}{K},\frac{{K\choose t+1}-{K-\min\{N,K\}\choose t+1}}{{K\choose t}}\bigg),\quad t\in[0:K].\label{eqn:achievable:LFR}
\end{IEEEeqnarray}
Moreover, the point $(0,N)$ can be achieved with subpacketization $1$, and the point $(M_t,R_t')$
can be achieved with subpacketization $F_t := {K\choose t}, \ t\in[0:K]$.
\end{theorem}}
 
For both Theorem~\ref{thm:1} and~\ref{thm:2},  for the point $(M,R)=(0,N)$, the server can trivially transmit all the $N$ files to the users, obviously this scheme satisfies both the correctness and the privacy condition.
For $t=K$, the result is trivial, since all users can cache all the files. For $t\in[0:K-1]$, we prove the theorems by analyzing the performances of the privacy key for SFR demands  in Section~\ref{sec:scheme} and privacy key for LFR  demands in Section~\ref{subsec:LFR}, respectively. The other points on the lower convex envelope can be achieved by memory-sharing between those points. 

%

\begin{table*}
\centering
\scalebox{1}{\begin{threeparttable}
\caption{Comparison of Different Schemes}\label{table:1}
\begin{tabular}{|c|c|c|c|c|c|}
  \hline
\tabincell{c}{ Scheme \\Demands} &\tabincell{c}{Non-private\\ SFR~\cite{Yu2019ExactTradeoff}, LFR~\cite{Kai2020LinearFunction}}  &\tabincell{c}{Virtual users\\   SFR~\cite{Kamath2019}} &\tabincell{c}{Privacy key\\ SFR$^\dagger$} &\tabincell{c}{Privacy key\\ LFR$^\dagger$}\\ \hline
    $t$&$0\leq t\leq K$&$0\leq t\leq KN$&$0\leq t\leq K$&$0\leq t\leq K$\\\hline
   $M$   &  $\frac{tN}{K}$&$\frac{t}{K}$ & $1+\frac{t(N-1)}{K}$ & $1+\frac{t(N-1)}{K}$  \\\hline
 $R$  & $\frac{{K\choose t+1}-{K-\min\{N,K\}\choose t+1}}{{K \choose t}}$   & $\frac{{KN\choose t+1}-{(K-1)N\choose t+1}}{{KN\choose t}}$& $\frac{{K\choose t+1}-{K-\min\{N-1,K\}\choose t+1}}{{K\choose t}}$  & $\frac{{K\choose t+1}-{K-\min\{N,K\}\choose t+1}}{{K\choose t}}$ \\\hline
$F$  & ${K\choose t}$   &${KN\choose t}$  & ${K\choose t}$ & ${K\choose t}$\\
  \hline
\end{tabular}
\begin{tablenotes}
\item[$\dagger$]
\footnotesize{In the privacy key schemes, the load-memory tradeoff curve  was obtained by taking the lower convex envelope of the points in this column and a trivial point $(M,R)=(0,N)$, which can be achieved with subpacketization $F=1$.}
\end{tablenotes}
\end{threeparttable}}
\end{table*}

\md{For clarity, in Table~\ref{table:1} we list the performance of the corner points of various schemes where, in term of an integer parameter $t$ whose range is given in the second row, with the memory size $M$ in the third row one achieves load $R$ and subpacketization $F$ given in the last two rows. 
If $N>K$ or $N\leq K, t>N-K$, the privacy key for SFR and LFR demands has the same load as the MAN scheme, given by 
\begin{IEEEeqnarray}{c}
R_t=R_t'=\frac{K-t}{t+1}. 
\end{IEEEeqnarray}  
If $N\leq K$ and $t\leq N-K$, the privacy key scheme for  LFR demands has slightly higher load than that for SFR demands. Thus, allowing the users to decode any linear combination of files slightly increase the load in this regime. Some numerical results are presented in Section \ref{sec:numerical}.}

\md{
\begin{remark}[Cost of Privacy]
We compare the performance of the non-private schemes with that of private schemes to quantify the cost of preserving demand privacy.
We have the following observations.
\begin{enumerate}

\item
The virtual user scheme has extremely high subpacketization, since it is based on a non-private scheme for $KN$ users. It also 
possibly suboptimal since (i) some multicast signals are only useful for the virtual users, and (ii) the multicast signals are from a non-private scheme designed for the case where all possible demands can occur, while in virtual user schemes the only demands possible are those where everyone of the $N$ files is requested exactly $K$ times.

\item 
The cache size of the privacy key scheme is larger than in the non-private scheme (by $1-\frac{t}{K}$), and the privacy is guaranteed with same or even better load (the load of the privacy key scheme for SFR demands is a function of $N-1$, while for non-private schemes it is a function of $N$). We will see that, the additional cache size is used to cache privacy keys. 

\item
Both the virtual user and the privacy key schemes need 
some randomness at the server, while no randomness is needed in the non-private case. 
It is an open question to characterize the minimum amount of randomness needed to guarantee privacy.
\end{enumerate}
\end{remark}
}
\md{
\begin{remark}[Memory Sharing and Subpacketization] The subpacketization of a coded caching scheme is the least number of packets each file needs to be split into, which is an important factor affecting the complexity of the scheme. 
In Theorem~\ref{thm:1} for SFR, or Theorem~\ref{thm:2} for LFR, for a general non-corner point $(M,R)$, assume that it is achieved by memory-sharing two corner points 
$(M_a,R_a)$ and $(M_b,R_b)$ with subpacketizations $F_a$ and $F_b$, respectively. That is, there exists a unique $\alpha\in(0,1)$ such that $M=\alpha M_a+(1-\alpha) M_b$ and $R=\alpha R_a+(1-\alpha) R_b$. To achieve the point $(M,R)$, each file $W_n, n\in[N],$ is partitioned into two subfiles $W_{n,a}$ and $W_{n,b}$, with sizes $\alpha B$  and $(1-\alpha) B$, respectively. The schemes achieving $(M_a,R_a)$ and $(M_b,R_b)$ are applied to the subfiles $\{W_{n,a}:n\in[N]\}$ and $\{W_{n,b}:n\in[N]\}$, respectively. Thus, the sub-packetization of the scheme achieving $M$ is given by $F_a+F_b$.
\end{remark}
}

Before we give the details of the general scheme, we illustrate the idea behind the privacy key schemes through an example.

\subsection{The privacy key scheme for the case $(N,K)=(3,2)$}
\label{sec:illustrative example}
Consider an $(N,K)=(3,2)$ DPCU  system with $t=1$.
Split each file into two equal-size packets, i.e., $W_1=\{W_{1,1},W_{1,2}\}$, $W_2=\{W_{2,1},W_{2,2}\}$ and $W_3=\{W_{3,1},W_{3,2}\}$.

We start with the case of SFR demands, \dt{in which case, without loss of generality, it suffices to consider $q=2$~\cite{Maddah-Ali2014fundamental}.}
In the placement phase, the server first generates two binary vectors $\bm{p}_1=(p_{1,1},p_{1,2},p_{1,3})^\mathsf{T}$ and $\bm{p}_2=(p_{2,1},p_{2,2},p_{2,3})^\mathsf{T}$ uniformly and independently at random from $\{(1,0,0)^\mathsf{T}, (0,1,0)^\mathsf{T}, (0,0,1)^\mathsf{T}, (1,1,1)^\mathsf{T}\}$, i.e., the set of binary vectors of length~3 whose Hamming weight is odd. Then the server generates two keys $S_1$ and $S_2$, one for each user, as follows
\begin{IEEEeqnarray}{rCl}
S_1&=&p_{1,1}W_{1,2}\oplus p_{1,2}W_{2,2}\oplus p_{1,3}W_{3,2},\quad\\
S_2&=&p_{2,1}W_{1,1}\oplus p_{2,2}W_{2,1}\oplus p_{2,3}W_{3,1}.
\end{IEEEeqnarray}
The contents of the caches are given by
\begin{IEEEeqnarray}{rCl}
Z_1&=&\{W_{1,1},W_{2,1},W_{3,1},S_1\},\label{eqn:Z1}\\
Z_2&=&\{W_{1,2},W_{2,2},W_{3,2},S_2\}.\label{eqn:Z2}
\end{IEEEeqnarray}

In the delivery phase, assume user $1$ demands $W_1$ and and user $2$ demands $W_2$, i.e., $\sd{d}_1=(1,0,0)^{\mathsf{T}}, \sd{d}_2=(0,1,0)^{\mathsf{T}}$.
In the non-private MAN scheme, the server sends the signal $W_{1,2}\oplus W_{2,1}$.  Here the server sends $X=(\sd{q}_1,\sd{q}_2,Y)$, where
\begin{subequations}\label{eqn:exam:signals}
\begin{IEEEeqnarray}{rCl}
\sd{q}_1&=&\sd{p}_1\oplus \sd{d}_1=(p_{1,1}\oplus 1,p_{1,2},p_{1,3})^{\mathsf{T}},\label{eqn:exam:q1}\\
\sd{q}_2&=&\sd{p}_2\oplus\sd{d}_2=(p_{2,1},p_{2,2}\oplus 1, p_{2,3})^{\mathsf{T}},\label{eqn:exam:q2}\\
Y&=&(W_{1,2}\oplus S_1)\oplus (W_{2,1}\oplus S_2).
\end{IEEEeqnarray}
\end{subequations}

 Notice that
 \begin{IEEEeqnarray}{rCl}
 W_{1,2}\oplus S_1&=&(p_{1,1}\oplus 1)W_{1,2}\oplus p_{1,2}W_{2,2}\oplus p_{1,3}W_{3,2},\label{eqn:oplus:1}\\
 W_{2,1}\oplus S_2&=&p_{2,1}W_{1,1}\oplus(p_{2,2}\oplus 1)W_{2,1}\oplus p_{2,3}W_{3,1}.\label{eqn:oplus:2}
 \end{IEEEeqnarray}
 By~\eqref{eqn:Z1} and~\eqref{eqn:oplus:1} (resp.~\eqref{eqn:Z2} and~\eqref{eqn:oplus:2}),  user $1$ (resp. $2$) can recover $W_{1,2}\oplus S_1$ (resp. $W_{2,1}\oplus S_2$) by computing and canceling $W_{2,1}\oplus S_2$ (resp. $W_{1,2}\oplus S_1$) from $Y$  using the vector  $\sd{q}_2$ (resp.  $\sd{q}_1$) and the contents of its cache. Hence user $1$ and $2$  can further decode their un-cached packet using their cached keys $S_1$ and $S_2$ respectively. The privacy is guaranteed since each user does not know the key of the other user, and the vectors $\sd{q}_1,\sd{q}_2$ are uninformly and independently distributed over $\{(0,0,0)^\mathsf{T},(1,1,0)^\mathsf{T},(1,0,1)^\mathsf{T},(0,1,1)^\mathsf{T}\}$, i.e., the set of binary vectors of length~3 whose Hamming weight is even.

Notice that each user caches $4$ packets, each of size $\frac{B}{2}$ bits. In the signal $X$, the main payload $Y$ is a coded packet of length $\frac{B}{2}$ and \md{the vectors $\sd{q}_1,\sd{q}_2$ can be sent in 
 $6$ bits,} which does not scale with $B$. Thus, the scheme achieves the memory-load pair $(M,R)=\big(2,\frac{1}{2}\big)$.

\yq{
Next, we consider the case of LFR demands. 
 The demands $\sd{d}_1=(d_{1,1},d_{1,2},d_{1,3})^{\mathsf{T}}$ and $\sd{d}_2=(d_{2,1},d_{2,2},d_{2,3})^{\mathsf{T}}$ can be arbitrary vectors in $\mathbb{F}_2^3$. In this case, the placement phase is the same form as~\eqref{eqn:Z1} and~\eqref{eqn:Z2}, except that the vectors $\sd{p}_1,\sd{p}_2$ will be uniformly chosen from $\mathbb{F}_2^{3}$, \dt{i.e., we no longer restrict them to be form a linear subspace of $\mathbb{F}_2^{3}$.} 
The signal $X=(\sd{q}_1,\sd{q}_2,Y)$ is generated similarly to~\eqref{eqn:exam:signals} with $\sd{q}_k=\sd{p}_k\oplus\sd{d}_k$ for $k=1,2$, buț with 
\begin{IEEEeqnarray}{rCl}
Y&=&\Big(\bigoplus_{j\in[3]} d_{1,j} W_{j,2}\oplus S_1\Big)\oplus\Big(\bigoplus_{j\in[3]}d_{2,j}W_{j,1}\oplus S_2\Big)\\
&=&\Big(\bigoplus_{j\in[3]} (p_{1,j}\oplus d_{1,j}) W_{j,2}\Big)\oplus\Big(\sum_{j\in[3]}(p_{2,j}\oplus d_{2,j})W_{j,1}\Big).
\end{IEEEeqnarray}
Similarly to the SFR demands, each user can decode its demanded linear combination of the files. The privacy is guaranteed since the vectors $\sd{q}_1,\sd{q}_2$ are uniformly distributed over $\mathbb{F}_2^3$. \md{
Notice that the range of $\sd{p}_1,\sd{p}_2$ can not be restricted to the odd weight vectors as we do so, for example, user $2$ may deduce the weight of $\sd{d}_1$ from the weight of $\sd{q}_1$, which violates the privacy condition.}

\dt{Also in this case, as for the SFR demands, the proposed scheme achieves the memory-load pair $(M,R)=\big(2,\frac{1}{2}\big)$.}

}


\begin{remark}[Comparison with non-private schemes]
\label{remark:comp:MN}
It can be observed from the above example that, compared to the non-private MAN scheme~\cite{Maddah-Ali2014fundamental},  the file are partitioned in the same way. The placement phase is similar to MAN; in addition, each user also caches a random linear combination of the uncached packets under the MAN placement, which is used as a key. The placement is thus not uncoded.
In the delivery phase, the server broadcasts a coded signal so that each user can decode a linear combination of files as per the scheme in~\cite{Kai2020LinearFunction}. The linear combination is designed such that each user can decode its demanded file with its cached key. \md{The reason that we constrain the vectors $\sd{p}_1,\sd{p}_2$ for SFR demands is that in this way the query vectors  $\sd{q}_1,\sd{q}_2$ in~\eqref{eqn:exam:signals} constrained in an $N-1=2$ dimensional subspace of $\mathbb{F}_2^3$. It will be clear later that, this constrain for SFR  demands will slightly improve the achievable load over the LFR demands in some parameter regimes.}
\end{remark}


\subsection{Privacy Key Scheme for  SFR Demands}
\label{sec:scheme} 
For any $t\in[0:K]$, let
\begin{IEEEeqnarray}{c}
\mathbf{\Omega}_t := \{\mathcal{T}:\mathcal{T}\subseteq[K],\,|\mathcal{T}|=t\}.
\end{IEEEeqnarray}
For fixed $t\in[0:K-1]$, the system operates as follows.
The server partitions the  file $\sys{W}_n$ into ${K\choose t}$ equal-size packets denoted as
\begin{IEEEeqnarray}{c}
\sys{W}_{n}=\big\{\sys{W}_{n,\mathcal{T}}:\,\mathcal{T}\in\mathbf{\Omega}_t\big\},\quad\forall n\in[N].\label{eqn:packets}
\end{IEEEeqnarray}
For notational simplicity, for any $\sd{a}=(a_1,\ldots,a_N)^{\mathsf{T}}\in\mathbb{F}_q^N$, we will use the notation 
\begin{IEEEeqnarray}{c}
W_{\sd{a},\mathcal{T}} :=  \sum_{n\in[N]} a_n \cdot W_{n,\mathcal{T}}, \quad\forall \mathcal{T}\in\mathbf{\Omega}_t, \label{eqn:Wa:linear:packet}
\end{IEEEeqnarray}
to denote a linear combination of packets.

\paragraph*{Placement Phase} 
The server uniformly and independently generates $K$ vectors $\bm{p}_1,\ldots,\bm{p}_K$ from the set of  all vectors in $\mathbb{F}_q^N$  such that the sum of the components is $q-1$, i.e.,
\begin{IEEEeqnarray}{c}
\bm{p}_k := (p_{k,1},\ldots,p_{k,N})^\mathsf{T}\sim\textnormal{Unif}\Big\{(x_1,\ldots,x_N)^\mathsf{T}\in\mathbb{F}_q^N\,:\,\mathop{\sum}\limits_{n\in[N]}x_n=q-1\Big\},\quad \forall\,k\in[K].\IEEEeqnarraynumspace\label{u:dist}
\end{IEEEeqnarray}
The server fills the cache of user $k\in[K]$ as
\begin{subequations}\label{eqn:cache}
\begin{IEEEeqnarray}{rCl}
\sys{Z}_k&=&\big\{\sys{W}_{n,\mathcal{T}}:\mathcal{T}\in\mathbf{\Omega}_t,k\in\mathcal{T},n\in[N]\big\}\label{eqn:cache:a}\\
&&\cup\big\{W_{\sd{p}_k,\mathcal{T}}:\mathcal{T}\in\mathbf{\Omega}_t, k\notin\mathcal{T}\big\}.\label{eqn:cache:b}
\end{IEEEeqnarray}
\end{subequations}
The random variable $\sys{P}$ is given by $\sys{P} = \sd{p}_{[K]}$.

\paragraph*{Delivery Phase} 
After receiving the users' demands $\sd{d}_{[K]}$, the server generates $K$ vectors 
\begin{IEEEeqnarray}{c}
\sd{q}_k=\sd{p}_k+\sd{d}_k,\quad \forall\,k\in[K].\label{eqn:C}
\end{IEEEeqnarray}
Denote the rank of matrix composed by the vectors $\sd{q}_{[K]}$ over the field $\mathbb{F}_q$ by $\textnormal{rank}_q(\sd{q}_{[K]})$. Let $\mathcal{L}$ be a fixed subset of $[K]$ of size $\textnormal{rank}_q(\sd{q}_{[K]})$ such that the vectors $\sd{q}_{\mathcal{L}}$ 
are linearly independent. 
Define
\begin{IEEEeqnarray}{c}
\sys{Y}_{\mathcal{S}} := \mathop{\sum}\limits_{j\in\mathcal{S}}W_{\sd{q}_j,\mathcal{S}\backslash\{j\}},\quad \forall\, \mathcal{S}\in\mathbf{\Omega}_{t+1}.  \label{eqn:XS}
\end{IEEEeqnarray}
The server transmits the signal 
\begin{IEEEeqnarray}{rCl}
X&=&(\mathcal{L},\sd{q}_{[K]},\sys{Y}),\, \text{where} \label{eqn:X}
\\
\sys{Y} :&=& \big\{\sys{Y}_{\mathcal{S}}:\mathcal{S}\in\mathbf{\Omega}_{t+1},\mathcal{S}\cap \mathcal{L}\neq\emptyset\big\}.\label{eqn:Y}
\end{IEEEeqnarray}
\md{
Note that the vectors $\sd{p}_{[K]}$ are designed such that the vectors $\sd{q}_{[K]}$ are uniformly and independently distributed over the $N-1$ dimensional subspace 
$
\big\{(x_1,\ldots,x_N)\in\mathbb{F}_q^N:\sum_{n\in[N]}x_n=0\big\}
$ by~\eqref{u:dist}. Thus, $\textnormal{rank}_q(\sd{q}_{[K]}) \leq \min\{N-1,K\}$.
}


\paragraph*{Proof of Correctness} By~\eqref{eqn:cache:a}, each user $k\in[K]$ needs to decode its demanded packets that were not cached.
For each packet $\sys{W}_{\sd{d}_k,\mathcal{T}}$ such that $k\notin\mathcal{T}$, user $k$ can decode $\sys{W}_{\sd{d}_k,\mathcal{T}}$ from the signal $\sys{Y}_{\mathcal{T}\cup\{k\}}$, since by~\eqref{eqn:C} and~\eqref{eqn:XS}, 
\begin{IEEEeqnarray}{c}
Y_{\mathcal{T}\cup\{k\}}=W_{\sd{d}_k,\mathcal{T}}+ W_{\sd{p}_k,\mathcal{T}}+\sum_{j\in\mathcal{T}}W_{\sd{q}_j,\mathcal{T}\cup\{k\}\backslash\{j\}}.
\end{IEEEeqnarray}
Notice that, $W_{\sd{p}_k,\mathcal{T}}$ is cached by user $k$ from~\eqref{eqn:cache:b}, and user $k$ can compute all the coded packets $W_{\sd{q}_j,\mathcal{T}\cup\{k\}\backslash\{j\}}$ for $j\in\mathcal{T}$ since $k\in\mathcal{T}\cup\{k\}\backslash\{j\}$ and  user $k$ knows the coefficient vectors $\sd{q}_{[K]}$  from the signal $X= (\mathcal{L},\sd{q}_{[K]},\sys{Y})$.

%

We still need to prove that each user can obtain all the signals
\begin{IEEEeqnarray}{c}
\{\sys{Y}_\mathcal{S}\,:\,\mathcal{S}\in\mathbf{\Omega}_{t+1},\mathcal{S}\cap\mathcal{L}=0\},\label{eqn:allYS}
\end{IEEEeqnarray}
which are not included in $Y$ (see~\eqref{eqn:Y}).
We note that the signals $\sys{Y}_{\mathcal{S}}$ in~\eqref{eqn:XS} in the main payload are exactly the same as in the non-private case where each user $k$ demands  the linear combination of  files~$W_{\sd{q}_k}$~\cite{Kai2020LinearFunction}.
It has been proved in~\cite{Kai2020LinearFunction} that the signals in~\eqref{eqn:allYS} can be obtained by linear combinations of those in~\eqref{eqn:Y}.

%

\paragraph*{Proof of Privacy}\label{subsec:privacy}
We prove the scheme satisfy the equivalent condition in~\eqref{eqn:privacyII}. In fact, for any $\mathcal{S}\subseteq[K]$,
\begin{subequations}
\begin{IEEEeqnarray}{rCl}
&&I(\sd{d}_{[K]\backslash\mathcal{S}};\sys{X},\sd{d}_\mathcal{S},\sys{Z}_\mathcal{S}, \sys{W}_{[N]})\\
&=&I(\sd{d}_{[K]\backslash\mathcal{S}};\mathcal{L},\sd{q}_{[K]},\sys{Y},\sd{d}_\mathcal{S},\sys{Z}_\mathcal{S}\,\sys{W}_{[N]})\\
&\stackrel{(a)}{=}&I(\sd{d}_{[K]\backslash\mathcal{S}};\mathcal{L},\sd{q}_{[K]},\sd{d}_\mathcal{S},\sys{Z}_\mathcal{S},\sys{W}_{[N]})\\
&\stackrel{(b)}{=}&I(\sd{d}_{[K]\backslash\mathcal{S}};\sd{q}_{[K]},\sd{d}_\mathcal{S},\sys{Z}_\mathcal{S},\sys{W}_{[N]})\\
&\leq&I(\sd{d}_{[K]\backslash\mathcal{S}};\sd{q}_{[K]},\sd{d}_\mathcal{S},\sys{Z}_\mathcal{S},\bm{p}_\mathcal{S},\sys{W}_{[N]})\\
&\stackrel{(c)}{=}&I(\sd{d}_{[K]\backslash\mathcal{S}};\sd{q}_{[K]\backslash\mathcal{S}},\sd{d}_\mathcal{S},\bm{p}_\mathcal{S},\sys{W}_{[N]})\\
&=&I(\sd{d}_{[K]\backslash\mathcal{S}};\sd{d}_\mathcal{S},\bm{p}_\mathcal{S},\sys{W}_{[N]})
  +I(\sd{d}_{[K]\backslash\mathcal{S}};\sd{q}_{[K]\backslash\mathcal{S}}\,|\,\sd{d}_\mathcal{S},\bm{p}_\mathcal{S},\sys{W}_{[N]})\\
&=&I(\sd{d}_{[K]\backslash\mathcal{S}};\bm q_{[K]\backslash\mathcal{S}}\,|\,\sd{d}_\mathcal{S},\bm{p}_\mathcal{S},\sys{W}_{[N]})\\
&=&H(\bm q_{[K]\backslash\mathcal{S}}\,|\,\sd{d}_\mathcal{S},\bm{p}_\mathcal{S},\sys{W}_{[N]})-H(\bm q_{[K]\backslash\mathcal{S}}\,|\,\sd{d}_{[K]},\bm{p}_\mathcal{S},\sys{W}_{[N]})\\
&\stackrel{(d)}{=}&H(\bm q_{[K]\backslash\mathcal{S}})-H(\bm p_{[K]\backslash\mathcal{S}}\,|\,\sd{d}_{[K]},\bm{p}_\mathcal{S},\sys{W}_{[N]})\label{eqn:explain}\\
&=&H(\bm q_{[K]\backslash\mathcal{S}})-H(\bm p_{[K]\backslash\mathcal{S}})\\
&\stackrel{(e)}{=}&0,\label{eqn:zero}
\end{IEEEeqnarray}
\end{subequations}
\yq{
where $(a)$ holds since by~\eqref{eqn:XS} and~\eqref{eqn:Y}, and $Y$ is determined by $\mathcal{L},\sd{q}_{[K]},W_{[N]}$; $(b)$ holds since $\mathcal{L}$ is determined by $\sd{q}_{[K]}$; $(c)$ holds since $Z_\mathcal{S}$ is determined by $W_{[N]}$ and $\sd{p}_{\mathcal{S}}$ by~\eqref{eqn:cache}, and $\sd{q}_{\mathcal{S}}$ is determined by $\sd{p}_{\mathcal{S}}$ and $\sd{d}_{\mathcal{S}}$ by~\eqref{eqn:C}; $(d)$ holds because $\bm q_{[K]\backslash\mathcal{S}}$ are independent of $(\sd{d}_\mathcal{S},\bm{p}_\mathcal{S},\sys{W}_{[N]})$ and $\bm q_{[K]\backslash\mathcal{S}}$ and $\bm p_{[K]\backslash\mathcal{S}}$ determines each other given $\sd{d}_{[K]\backslash \mathcal{S}}$; and 
$(e)$ holds because $\bm q_{[K]}$ are uniformly distributed over an $N-1$ dimension subspace of $\mathbb{F}_q^N$ and $\bm p_{[K]}$ are uniformly distributed over the coset of an $N-1$ dimensional subspace of $\mathbb{F}_q^N$ in~\eqref{u:dist} by construction.
}

\paragraph*{Performance}
By~\eqref{eqn:packets}, each file is equally split into ${K\choose t}$ packets, each of size $B / {K\choose t}$, thus the subpacketization is ${K\choose t}$. Moreover, by~\eqref{eqn:cache}, the number of packets cached by each user is $N{K-1\choose t-1}+{K-1\choose t}$, thus the  memory size is
 \begin{IEEEeqnarray}{rCl}
 M_t&=&\frac{1}{B}\cdot \left(N{K-1\choose t-1}+{K-1\choose t}\right)\cdot \frac{B}{{K\choose t}}
=\frac{K+t(N-1)}{K}.
 \end{IEEEeqnarray}
By~\eqref{eqn:XS} and~\eqref{eqn:Y}, the main payload $\sys{Y}$ contains ${K\choose t+1}-{K-\textnormal{rank}_q{(\sd{q}_{[K]})}\choose t+1}$ packets, each of size $B / {K\choose t}$.  Notice that the set $\mathcal{L}$ and the vectors $\sd{q}_{[K]}$   can be sent in $K$ and $NK$ symbols, respectively. By~\eqref{u:dist} and~\eqref{eqn:C}, $\bm{q}_1,\ldots,\bm{q}_{K}$ are uniformly and independently distributed over
 the $N-1$ dimensional subspace $
\big\{(x_1,\ldots,x_N)\in\mathbb{F}_q^N:\sum_{n\in[N]}x_n=0\big\}
$ of $\mathbb{F}_q^N$. Thus, the worst-case is $\max \textnormal{rank}_q(\sd{q}_{[K]})=\min\{N-1,K\}$. Therefore, the scheme achieves the worst-case load
 \begin{IEEEeqnarray}{rCl}
 R_t&=&\liminf_{B\rightarrow\infty}\frac{1}{B}\bigg(\bigg({K\choose t+1}-{K-\min\{N-1,K\}\choose t+1}\bigg)\cdot\frac{B}{{K\choose t}}+K+NK\bigg)
 \notag\\
 &=&
\frac{{K\choose t+1}-{K-\min\{N-1,K\}\choose t+1}}{{K\choose t}}.\label{eqn:Rt}
 \end{IEEEeqnarray}

This concludes the description of the general  privacy key for SFR demands.

\subsection{Privacy Key  Scheme for LFR Demands}\label{subsec:LFR}


\yq{
The above privacy key  for SFR demands can be adapted to LFR demands as follows. The range of the independent and uniformly distributed vectors $\bm p_1,\ldots,\bm p_K$ in~\eqref{u:dist} is now $\mathbb{F}_q^N$. 
 In this case, we can not constrain them to something equivalent to a coset of a subspace as in~\eqref{u:dist} because the demand $\sd{d}_k$ may takes all vectors over $\mathbb{F}_q^N$, so the range of $\sd{q}_k$ ($k\in[K]$) needs at least $|\mathbb{F}_q^N|=q^N$ distinct vectors. 
}

The correctness and privacy can be verified following the same lines of the above proofs. The performance analysis follows the same lines except that the largest rank of the vectors $\sd{q}_{[K]}$ is given by $\max \textnormal{rank}_q(\sd{q}_{[K]})=\min\{N,K\}$ since  the vectors $\sd{q}_{[K]}$ are independent and uniformly distributed over $\mathbb{F}_q^N$ in this case, thus the worst-case load  in~\eqref{eqn:Rt} is replaced by
\begin{IEEEeqnarray}{c}
R_t' :=  \frac{{K\choose t+1}-{K-\min\{N,K\}\choose t+1}}{{K\choose t}}.\label{eqn:Rt:prime}
\end{IEEEeqnarray}

\md{
\begin{remark}[Relationship to PIR schemes] 
In Private Information Retrieval (PIR), a well known technique to decode a demanded packet from multiple servers while keeping the index of the file the packet belongs to private is to download two or more linear combinations of packets of different  files; each linear combination is from a distinct server and such that the packet to be decoded dominates an independent $1$-dimensional subspace of the subspace spanned by the linear combinations. The index of the demanded file is kept private from any individual server since the server has access only to one of the linear combinations downloaded by the user. We use a similar idea in our proposed privacy key scheme. With the privacy keys cached by each user $k\in[K]$ in~\eqref{eqn:cache:b} and the uncoded part~\eqref{eqn:cache:a},  user $k$ has access to a linear combination of the files $W_{[N]}$ with coefficient vector $\bm p_k$.  In the delivery phase,  user $k$ decodes another linear combination with coefficient vector $\bm q_k$, and the demanded message $W_{\sd{d}_k}$ dominates an independent subspace of the linear space spanned by the two linear combinations. The demand vector $\sd{d}_k$
 is kept private from all the other users since they know neither the other linear combination, nor the coefficient vector $\bm p_k$.  
\end{remark}
}

\section{Converse and Optimality}
\label{sec:converse}
\md{In this section, we derive a converse bound for DPCU with SFR demands.}
The converse is inspired by the proof in~\cite{Sneha2019}, which is summarized in Section~\ref{exam:converse} for the case $K=N=2$.
Our new converse is proved Section~\ref{sec:converse2}.
In Section~\ref{sec:opt} we prove that our privacy key scheme is order optimal.

\yq{
 For notational simplicity, in this section use the scalar variables $D_1,\ldots,D_K$ to denote the index of the demanded files by the users, that is, $D_k=n$ is equivalent to $\sd{d}_k=\bm{e}_n$ for all $k\in[K],n\in[N]$.
 }
In order to establish our converse,  we use the submodularity of entropy functional as in the following lemma.
\begin{lemma}[Submodularity of Entropy~\cite{submodular_book}]
Let $\mathcal{X}$ be a set of random variables. For any $\mathcal{X}_1,\mathcal{X}_2\subseteq\mathcal{X}$,
\begin{IEEEeqnarray}{c}
H(\mathcal{X}_1)+H(\mathcal{X}_2)\geq H(\mathcal{X}_1\cup\mathcal{X}_2)+H(\mathcal{X}_1\cap\mathcal{X}_2)\label{eqn:submodualrity}.
\end{IEEEeqnarray}
\end{lemma}

\md{
\subsection{Example of converse for the case $N=K=2$~\cite{Sneha2019}}
\label{exam:converse}
Consider the case $N=K=2$. 
If user $k\in[2]$ demands $D_k=n\in[2]$, by the correctness condition the file $W_n$ must be decodable from $(X,Z_k)$, that is
\begin{subequations}
\begin{IEEEeqnarray}{rCl}
B &=& H(W_n) 
 \stackrel{(a)}{=} H(W_n\,|\,D_k=n)  \\
&\leq& H(W_n,X,Z_k\,|\,D_k=n)\label{eq:exam:leftside}\\
&\stackrel{(b)}{=}&H(X,Z_k\,|\,D_k=n)\\
&\leq& H(X\,|\,D_k=n)+H(Z_k\,|\,D_k=n)\\
&\stackrel{(c)}{=}& H(X)+H(Z_k)
\leq (R+M)B, \label{eqn:RM} 
\end{IEEEeqnarray}
\label{eqn:RMall}
\end{subequations}
where the inequalities in~\eqref{eqn:RMall} follow from:
$(a)$ from the independence condition in~\eqref{eqn:indep.cond.}, i.e., $W_n$ is independent of $D_k$;
$(b)$ from the correctness condition in~\eqref{eqn:correctness}; and 
$(c)$ from the privacy condition in~\eqref{eqn:privacyII}, i.e., $X$ is independent of $D_k$, 
 and the fact that $Z_j=\varphi_k(\sys{P},\sys{W}_{[N]})$ as defined in~\eqref{eqn:Zk}  is independent of $D_k$ 
 since $(\sys{P},\sys{W}_{[N]})$  is independent of $D_k$ from the independence condition in~\eqref{eqn:indep.cond.}.

Thus, at this point, we have $R+M\geq 1$ from~\eqref{eqn:RMall}, which we further improve as follows.
\begin{itemize}

\item
Step~1: Combine two terms of the form $H(W_1,X,Z_j\,|\,D_j=1)$ via submodularity:
\begin{subequations}
\begin{IEEEeqnarray}{rCl}
&&H(W_1,X,Z_2\,|\,D_2=1)+H(W_1,X,Z_1|\,D_1=1)\\
&\stackrel{(a)}{=}&H(W_1,X,Z_2\,|\,D_1=1,D_2=1)+H(W_1,X,Z_1\,|\,D_1=1,D_2=1)\\
&\stackrel{(b)}{\geq}&H(W_1,X\,|\,D_1=1,D_2=1)+H(W_1,X,Z_1,Z_2\,|\,D_1=1,D_2=1)\\
&\geq&H(W_1,X\,|\,D_1=1,D_2=1)+H(W_1,Z_1,Z_2\,|\,D_1=1,D_2=1)\\
&\stackrel{(c)}{=}&H(W_1,X)+H(W_1,Z_1,Z_2)\label{eq:exam:lower}
=: h(1),
\end{IEEEeqnarray}
\label{eqn:all123}
\end{subequations}
where the inequalities in~\eqref{eqn:all123} follow from:
$(a)$  from~\eqref{eqn:privacyII}; 
$(b)$ from~\eqref{eqn:submodualrity}; and 
$(a)$ from independence, i.e.,~\eqref{eqn:privacyII},~\eqref{eqn:Zk}, and~\eqref{eqn:indep.cond.}. 
For notation convenience we denote~\eqref{eq:exam:lower} by $h(1)$ to stress the fact that it involves the first file $W_1$.

\item
Step~2: Add one more term $H(W_1,X,Z_1\,|\,D_1=1)$ to~\eqref{eq:exam:lower}, and combine the three terms via submodularity:
\begin{subequations}
\begin{IEEEeqnarray}{rCl}
&& 
\dt{h(1)}+H(W_1,X,Z_1\,|\,D_1=1)\label{exam:first:step}\\
&\stackrel{(a)}{=}&H(W_1,X)+(H(W_1,Z_1,Z_2\,|\,D_1=1,D_2=2)+H(W_1,X,Z_1\,|\,D_1=1,D_2=2))\\
&\stackrel{(b)}{\geq}&H(W_1,X)+H(W_1,Z_1\,|\,D_1=1,D_2=2)+H(W_1,X,Z_1,Z_2\,|\,D_1=1,D_2=2)\\
&\stackrel{(c)}{\geq}&H(W_1,X\,|\,D_1=2)+H(W_1,Z_1\,|\,D_1=2)+H(W_1,W_2,X,Z_1,Z_2\,|\,D_1=1,D_2=2)\IEEEeqnarraynumspace\\
&\stackrel{(d)}{\geq}&H(W_1,X, Z_1\,|\,D_1=2)+H(W_1\,|\,D_1=2)+H(W_1,W_2,Z_1,Z_2\,|\,D_1=1,D_2=2)\\
&\stackrel{(e)}{=}&H(W_1,W_2,X, Z_1\,|\,D_1=2)+H(W_1)+H(W_1,W_2,Z_1,Z_2)\\
&\geq&H(W_1,W_2,X\,|\,D_1=2)+H(W_1)+H(W_1,W_2,Z_1,Z_2)\\
&\stackrel{(f)}{=}&H(W_1,W_2,X)+H(W_1,W_2,Z_1,Z_2)+H(W_1)\label{exam:explan:step}\\
&=:&h(2)+H(W_1) \geq 
5B,\label{eqn:exam:5B}
\end{IEEEeqnarray}
\label{eqn:all456}
\end{subequations}
where the inequalities in~\eqref{eqn:all456} follow from:
$(a)$ holds due to~\eqref{eqn:Zk},~\eqref{eqn:indep.cond.}, and~\eqref{eqn:privacyII}; 
$(b)$ and $(d)$ from submodularity in~\eqref{eqn:submodualrity}; 
$(c)$ and $(e)$ from~\eqref{eqn:Zk},~\eqref{eqn:indep.cond.}, and the correctness condition~\eqref{eqn:correctness}; and 
$(f)$ from~\eqref{eqn:privacyII}.
Note that in~\eqref{exam:explan:step} we obtained the term $h(2)$ that involves the first two files $W_{[2]}=(W_1,W_2)$.

\end{itemize}
Therefore, by~\eqref{eqn:RM},~\eqref{eq:exam:lower} and~\eqref{eqn:exam:5B}, we obtain $3R+3M\geq 5$ and hence the lower bound of $R+M$ is improved from $1$ to $\frac{5}{3}$.

\begin{remark}[An Observation] \label{remark:observation} 
Our general converse is based on the following observation from the above proof for the case $N=K=2$. 
The two entropy terms in the function
\begin{IEEEeqnarray}{c}
h(a) :=  H(W_{[a]},X)+H(W_{[a]},Z_1,Z_2),\quad a\in[2],\label{exam:h}
\end{IEEEeqnarray} 
can be regarded as two ``boxes'' of decoded files, each containing the first $a$ files $W_{[a]}=(W_1,\ldots,W_a)$. 
Compare~\eqref{exam:first:step} and~\eqref{eqn:exam:5B}: 
by combining a new entropy term $H(W_1,X,Z_1\,|\,D_1=1)$ with the two ``boxes" in $h(1)$ via submodularity, one can increase the number of decoded files in each ``box'' by one, while the file $W_1$ in the new term $H(W_1,X,Z_1\,|\,D_1=1)$ can be kept in the last term in~\eqref{eqn:exam:5B}. 
Our key idea is that we can recursively combine new terms of the form $H(W_1,X,Z_1\,|\,D_1=1)$ via submodularity until the two ``boxes'' are full  (i.e., $a=N$ in~\eqref{exam:h}, here $N=2$). 
\end{remark}

\subsection{New Converse Bound}
\label{sec:converse2}


Our new converse is a generalization of the example in Section~\ref{exam:converse} from~\cite{Sneha2019} based on the ``induction'' observation in Remark~\ref{remark:observation}. In Lemma~\ref{lemma:conlusion:bound}, we generalize the bound in~\eqref{eqn:RMall} to obtain an initial lower bound for $R+\ell \cdot M$ for any $\ell\in[N]$. In Lemma~\ref{lemma:h}, we generalize the definition of $h(a)$ in~\eqref{exam:h} and formalize the rules of combining used in~\eqref{eqn:all456}. 
In Theorem~\ref{thm:lowerbound}, we first obtain a bound similar to~\eqref{eq:exam:lower}, and then apply the combining rule in Lemma~\ref{lemma:h} recursively to obtain an improved lower bound for $R+\ell\cdot M$ for any $\ell\in[N]$.
}

\begin{lemma}\label{lemma:conlusion:bound}
For an $(N,K)$ DPCU system, assume $(M,R)\in\mathbb{R}_+^2$ is achievable for SFR demands,  then for any $\ell\in[N]$ and $b\in[0:\min\{\ell,K\}]$, and for sufficiently large $B$, the following holds
\begin{IEEEeqnarray}{c}
(R+\ell\cdot  M)B\geq H(W_{\mathcal{B}_{\ell}},X,Z_{\mathcal{A}_b}\,|\,\{D_j=d_j\}_{j\in\mathcal{A}_b})\label{eqn:basic}
\end{IEEEeqnarray}
for any $\mathcal{B}_\ell\subseteq[N],\mathcal{A}_b\subseteq [K]$ such that $|\mathcal{B}_\ell|=\ell,|\mathcal{A}_b|=b$, and 
$\{d_j:j\in\mathcal{A}_b\}$  is a set of any  $b$ distinct indices in $\mathcal{B}_\ell$.
\end{lemma}

\begin{lemma}\label{lemma:h} For fixed $\ell\in[N-1]$, define $b_\ell := \min\{\ell,K-1\}$, which satisfies $b_\ell\leq \min\{\ell,K\}$ and $b_\ell+1=\min\{\ell+1,K\}\leq K$.
For any $a\in[\ell:N]$ define
\begin{IEEEeqnarray}{c}
h_\ell(a) :=  \sum_{j=0}^{b_\ell-1}H(W_{[a]},X,Z_{[j]}\,|\,D_{[j]}=[j])+H(W_{[a]},Z_{[b_\ell+1]}).\label{eqn:hla}
\end{IEEEeqnarray}
Then  for any $a\in[\ell:N-1]$, the following holds
\begin{IEEEeqnarray}{rCl}
h_{\ell}(a)+H(W_{[\ell]},X,Z_{[b_\ell]}\,|\,D_{[b_\ell]}=[b_\ell])\geq h_\ell(a+1)+H(W_{[\ell]}).\label{recursive:ineq}
\end{IEEEeqnarray}
\end{lemma}

We are now ready to present our new converse result.

\begin{theorem}\label{thm:lowerbound} For an $(N,K)$ DPCU system, for any $M\in[0,N]$, the optimal memory-load tradeoff for SFR demands satisfies 
\begin{IEEEeqnarray}{c}
R_{\rm{F}}^*(M)\geq \max_{\ell\in[N]}\Big\{ \ell+\frac{\min\{\ell+1,K\} \ (N-\ell)}{N-\ell+\min\{\ell+1,K\}}-\ell \ M\Big\}.\label{eqn:lower:bound}
\end{IEEEeqnarray}
\end{theorem}

\begin{IEEEproof} Notice that, for fixed $\ell\in[N-1]$ and for sufficiently large $B$, we have
\begin{subequations}
\begin{IEEEeqnarray}{rCl}
&&(N-\ell+b_\ell+1)(R+\ell M)B\\
&=&b_\ell(R+\ell M)B+(N-\ell+1)(R+\ell M)B\\
&\stackrel{(a)}{\geq}&\sum_{j=0}^{b_\ell-2}H(W_{[\ell]},X,Z_{[j]}\,|\,D_{[j]}=[j]) +H(W_{[\ell]},X,Z_{[b_\ell-1]},Z_{b_\ell+1}\,|\,D_{[b_\ell-1]}=[b_\ell-1],D_{b_{\ell}+1}=\ell)\notag\\
&&+(N-\ell+1)H(W_{[\ell]},X,Z_{[b_\ell]}\,|\,D_{[b_\ell]}=[b_\ell])\label{ineq:A}\\
&\stackrel{(b)}{=}&\sum_{j=0}^{b_\ell-2}H(W_{[\ell]},X,Z_{[j]}\,|\,D_{[j]}=[j])+H(W_{[\ell]},X,Z_{[b_\ell-1]},Z_{b_\ell+1}\,|\,D_{[b_\ell]}=[b_\ell],D_{b_{\ell}+1}=\ell)\notag\\
&&+H(W_{[\ell]},X,Z_{[b_\ell]}\,|\,D_{[b_\ell]}=[b_\ell],D_{b_\ell+1}=\ell)+(N-\ell)H(W_{[\ell]},X,Z_{[b_\ell]}\,|\,D_{[b_\ell]}=[b_\ell])\IEEEeqnarraynumspace\label{eq:exp:B}\\
&\stackrel{(c)}{\geq}&\sum_{j=0}^{b_\ell-2}H(W_{[\ell]},X,Z_{[j]}\,|\,D_{[j]}=[j])+H(W_{[\ell]},X,Z_{[b_\ell-1]}\,|\,D_{[b_\ell]}=[b_\ell],D_{b_{\ell}+1}=\ell)\notag\\
&&+H(W_{[\ell]},X,Z_{[b_\ell+1]}\,|\,D_{[b_\ell]}=[b_\ell],D_{b_\ell+1}=\ell)+(N-\ell)H(W_{[\ell]},X,Z_{[b_\ell]}\,|\,D_{[b_\ell]}=[b_\ell])\IEEEeqnarraynumspace\\
&\geq&\sum_{j=0}^{b_\ell-2}H(W_{[\ell]},X,Z_{[j]}\,|\,D_{[j]}=[j])+H(W_{[\ell]},X,Z_{[b_\ell-1]}\,|\,D_{[b_\ell]}=[b_\ell],D_{b_{\ell}+1}=\ell)\notag\\
&&+H(W_{[\ell]},Z_{[b_\ell+1]}\,|\,D_{[b_\ell]}=[b_\ell],D_{b_\ell+1}=\ell)+(N-\ell)H(W_{[\ell]},X,Z_{[b_\ell]}\,|\,D_{[b_\ell]}=[b_\ell])\\
&\stackrel{(d)}{=}&\sum_{j=0}^{b_\ell-2}H(W_{[\ell]},X,Z_{[j]}\,|\,D_{[j]}=[j])+H(W_{[\ell]},X,Z_{[b_\ell-1]}\,|\,D_{[b_\ell-1]}=[b_\ell-1])\notag\\
&&+H(W_{[\ell]},Z_{[b_\ell+1]})+(N-\ell)H(W_{[\ell]},X,Z_{[b_\ell]}\,|\,D_{[b_\ell]}=[b_\ell])\\
&\stackrel{(e)}{=}&h_\ell(\ell)+(N-\ell)H(W_{[\ell]},X,Z_{[b_\ell]}\,|\,D_{[b_\ell]}=[b_\ell])\\
&\stackrel{(f)}{\geq}&h_\ell(\ell+1)+(N-\ell-1)H(W_{[\ell]},X,Z_{[b_\ell]}\,|\,D_{[b_\ell]}=[b_\ell])+H(W_{[\ell]})\label{eqn:rec1}\\
&&\quad\vdots\notag\\
&\stackrel{(g)}{\geq}&h_\ell(N)+(N-\ell)H(W_{[\ell]})\label{eqn:rec2}\\
&=&\sum_{j=0}^{b_\ell-1}H(W_{[N]},X,Z_{[j]}\,|\,D_{[j]}=[j])+H(W_{[N]},Z_{[b_\ell+1]})+(N-\ell)H(W_{[\ell]})\\
&\geq&\sum_{j=0}^{b_\ell-1}H(W_{[N]}\,|\,D_{[j]}=[j])+H(W_{[N]})+(N-\ell)H(W_{[\ell]})\\
&\stackrel{(h)}{=}&\sum_{j=0}^{b_\ell-1}H(W_{[N]})+H(W_{[N]})+(N-\ell)H(W_{[\ell]})\\
&=&(b_\ell+1)H(W_{[N]})+(N-\ell)H(W_{[\ell]})\label{eqn:last:step}
\end{IEEEeqnarray}
\end{subequations}
\md{
where we obtain $(a)$   by individually lower bounding  each $(R+\ell \cdot M)B$ by~\eqref{eqn:basic}, i.e., the following inequalities are used:
\begin{IEEEeqnarray}{rCl}
(R+\ell \cdot M)B&\geq& H(W_{[\ell]}, X,Z_{[j]}\,|\,D_{[j]}=[j]),\quad \forall\, j\in[0:b_{\ell}-2]\cup\{b_\ell\},\\
(R+\ell \cdot M)B&\geq&H(W_{[\ell]},X,Z_{[b_{\ell}-1]},Z_{b_{\ell}+1}\,|\,D_{[b_{\ell}-1]}=[b_{\ell}-1],D_{b_{\ell+1}}=\ell);
\end{IEEEeqnarray}
the equality $(b)$ holds since~\eqref{eqn:privacyII} implies that $D_{[K]\backslash\mathcal{A}_b}$ is independent of$(W_{[\ell]},X,Z_{\mathcal{A}_b},D_{\mathcal{A}_b})$\footnote{Notice that, for random variables $V_1,V_2,V_3$, the independence between the pair $(V_1,V_2)$ and $V_3$ implies  the independence of $V_1$ and $V_3$ conditioned  on any realization of $V_2$.} ; $(c)$ follows from the submodularity property~\eqref{eqn:submodualrity}; $(d)$ follows from~\eqref{eqn:Zk},~\eqref{eqn:indep.cond.} and~\eqref{eqn:privacyII}; $(e)$ follows from the definition in~\eqref{eqn:hla}; to get from~$(f)$ to~$(g)$, we recursively applied~\eqref{recursive:ineq} by setting $a=\ell,\ell+1,\ldots,N-1$; and $(h)$ follows from~\eqref{eqn:indep.cond.}. 
}

Finally, since the files are uniformly distributed over $\mathbb{F}_q^B$, we have $H(W_{[N]})=NB$
and thus 
\begin{IEEEeqnarray}{c}
R\geq \ell+\frac{\min\{\ell+1,K\}\cdot(N-\ell)}{N-\ell+\min\{\ell+1,K\}}-\ell\cdot  M\label{eqn:resultbound}
\end{IEEEeqnarray}
holds for $\ell\in[N-1]$. Moreover,  let $\ell=N$, $\mathcal{B}_{\ell}=[N]$ and $\mathcal{A}_b=\emptyset$ in~\eqref{eqn:basic} to obtain
\begin{IEEEeqnarray}{rCl}
(R+NM)B\geq H(W_{[N]},X)\geq H(W_{[N]})= NB.
\end{IEEEeqnarray}
Thus the inequality~\eqref{eqn:resultbound} also holds for $\ell=N$.
Therefore, we proved~\eqref{eqn:lower:bound}.
\end{IEEEproof}

\subsection{Optimality of the Privacy Key Scheme}
\label{sec:opt}
For clarity, we use $R_{\rm{F}}(M)$ and $R_{\rm{L}}(M)$ to denote the achievable load of the privacy key schemes for SFR demands in Section~\ref{sec:scheme} and LFR demands in Section~\ref{subsec:LFR}, respectively, i.e., the lower convex envelope of the points defined in Theorem~\ref{thm:1} and~\ref{thm:2}, respectively.

\dt{
The following theorem characterizes the optimality of $R_{\rm{F}}(M)$ and $R_{\rm{L}}(M)$ by leveraging the relationship in~\eqref{eq:TrivialRelationships}. 
A more refined result can be found in Appendix~\ref{sec:app:D} and in Appendix~\ref{App:sec:F} for $R_{\rm{F}}(M)$ and $R_{\rm{L}}(M)$, respectively.
\begin{theorem}\label{thm:onesinglegap}
From the relationship in~\eqref{eq:TrivialRelationships}, with $R_{\rm{F, converse}}(M)$ obtained from Theorem~\ref{thm:lowerbound} and other converse bounds without the privacy constraint and with $R_{\rm{L, achievable}}(M)$ from Theorem~\ref{thm:2}, for an $(N,K)$ DPCU system we have
\begin{IEEEeqnarray}{c}
\frac{R_{\rm{L, achievable}}(M)}{R_{\rm{F, converse}}(M)} \leq 6.3707,
\end{IEEEeqnarray}
that is, the privacy key scheme is optimal to within a constant gap in all parameter regimes.
\end{theorem}
}

\begin{remark}[Relations to known gap results]\label{remark:gap}  
It was showed in~\cite{Kamath2019} that the load achieved by the virtual users scheme for SFR demands in~\cite{Kamath2019} is optimal to within a multiplicative factor of $8$ if $N\leq K$, or of $4$ if $N>K$ and $M\geq\frac{N}{K}$, thus leaving open the regime $N>K, M<\frac{N}{K}$. Here, we show that the privacy key  for SFR demands is order optimal in all regimes under the the demand privacy condition~\eqref{eqn:privacy}, due to the new converse we derived in Theorem~\ref{thm:lowerbound}.  It was noticed in~\cite{KaiD2DPrivacy} that the virtual users scheme~\cite{Kamath2019} satisfies the privacy against user colluding for SFR demands. In fact, it satisfies our stronger privacy definition~\eqref{eqn:privacy}. Both privacy key  and virtual user schemes achieve the load $R=N$ at $M=0$, which is an optimal point under the new privacy definition in~\eqref{eqn:privacy}. 

Under the other privacy definitions in Remark~\ref{remark:privacy}, the best known converse bound at $M=0$ is $\min\{N,K\}$. Thus, the ratio between achievable and converse bounds for those privacy definitions is unbounded at $M=0$ when $N>K, M<\frac{N}{K}$. \md{It was showed in~\cite{Sneha2019} that the scheme described in Example~\ref{exam:1} achieves $(M,R)=\big(M,K(1-\frac{M}{N})\big)$, which is to within a multiplicative gap of $8$ for the regime $N>K, M<\frac{N}{K}$.  As observed in Example~\ref{exam:1}, the scheme does not satisfy our stronger privacy condition in~\eqref{eqn:privacy}.}

\end{remark}



\section{Performance Comparison and Numerical Results}\label{sec:numerical}

In this section, we compare the schemes in Table~\ref{table:1}. 
We compare the schemes in the two regimes $N\leq K$ and $N>K$, where we choose parameters $(N,K)=(10,30)$ and $(N,K)=(30,10)$. For both cases, we plot two figures showing:
\begin{enumerate}
  \item [(a)] the load-memory tradeoff curves of the schemes and the lower bounds in Theorem~\ref{thm:lowerbound} and~\cite{QYu2018Factor2};
  \item [(b)]  the subpacketization $F$ as a function of memory size $M$ for the corner points.
\end{enumerate}

\begin{figure}[htbp] \centering
\subfigure[Load-memory tradeoff] {
 \label{fig1:a}
\includegraphics[width=0.95\columnwidth]{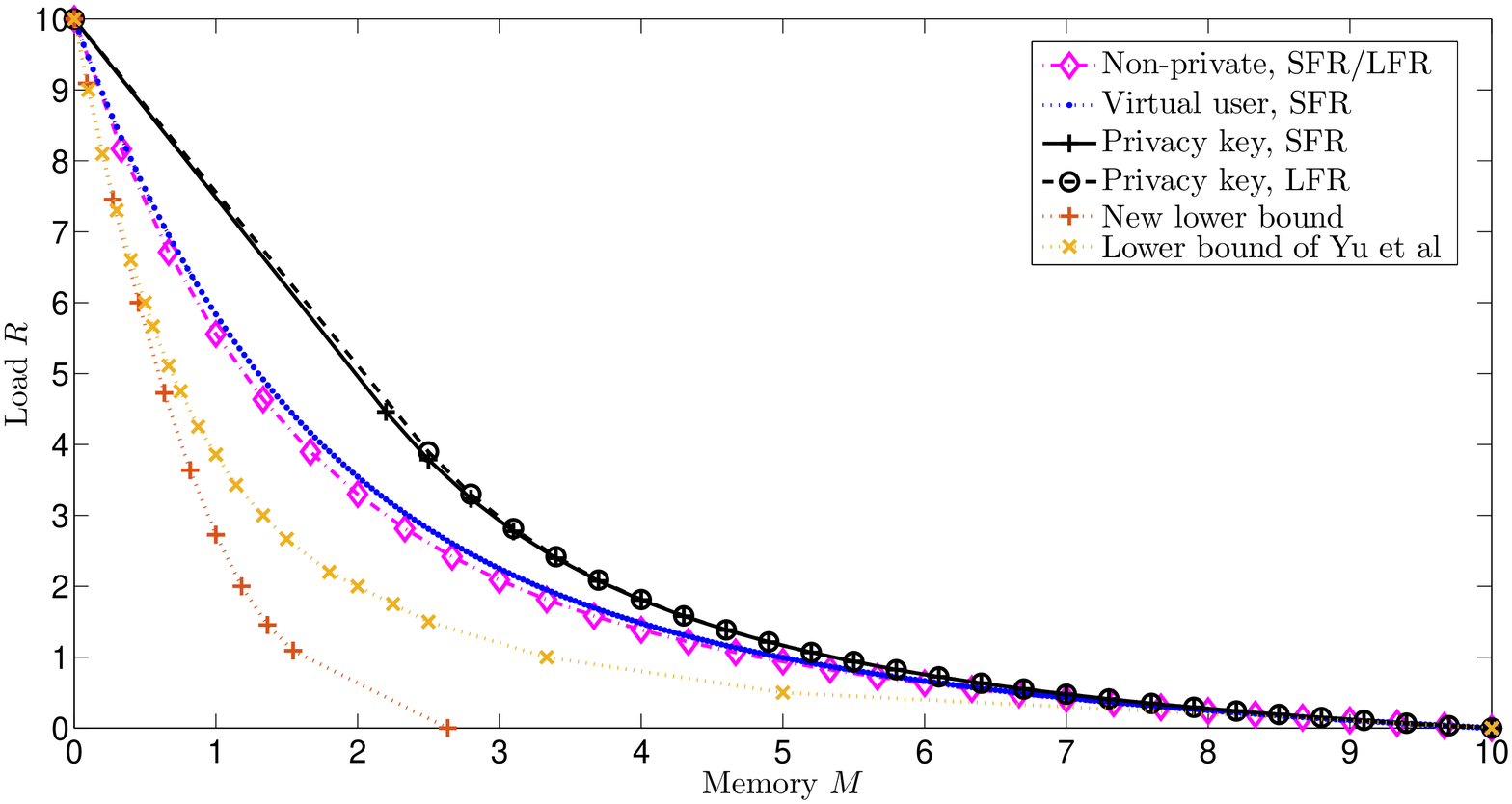}
}
\subfigure[ Subpacketization v.s. memory size] {
\label{fig1:b}
\includegraphics[width=0.95\columnwidth]{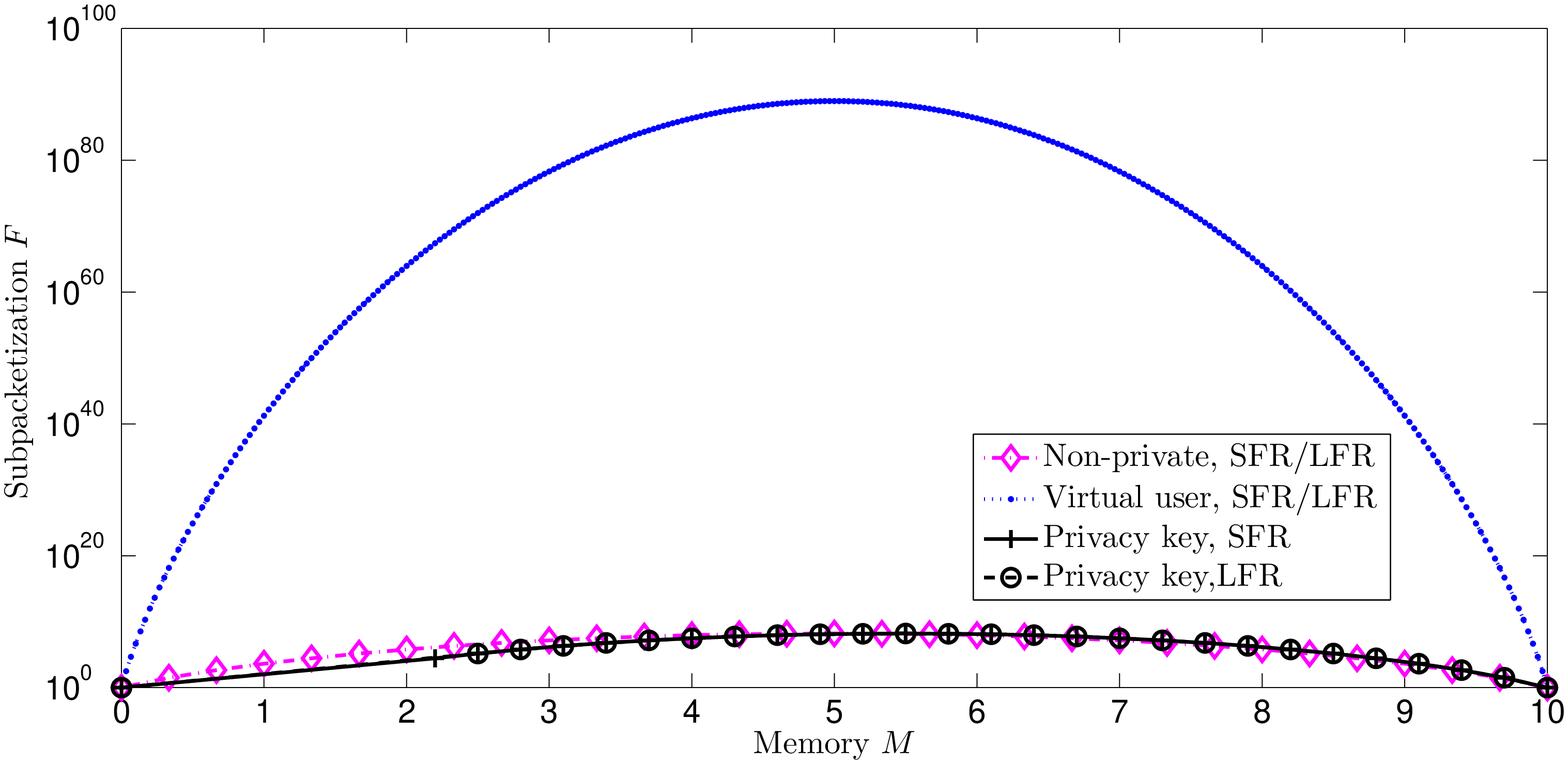}
}

\caption{Performance comparison for an $(N,K)=(10,30)$ system.  }
\label{fig:1}
\end{figure}

\begin{figure}[htbp] \centering
\subfigure[Load-memory tradeoff] {
 \label{fig2:a}
\includegraphics[width=0.95\columnwidth]{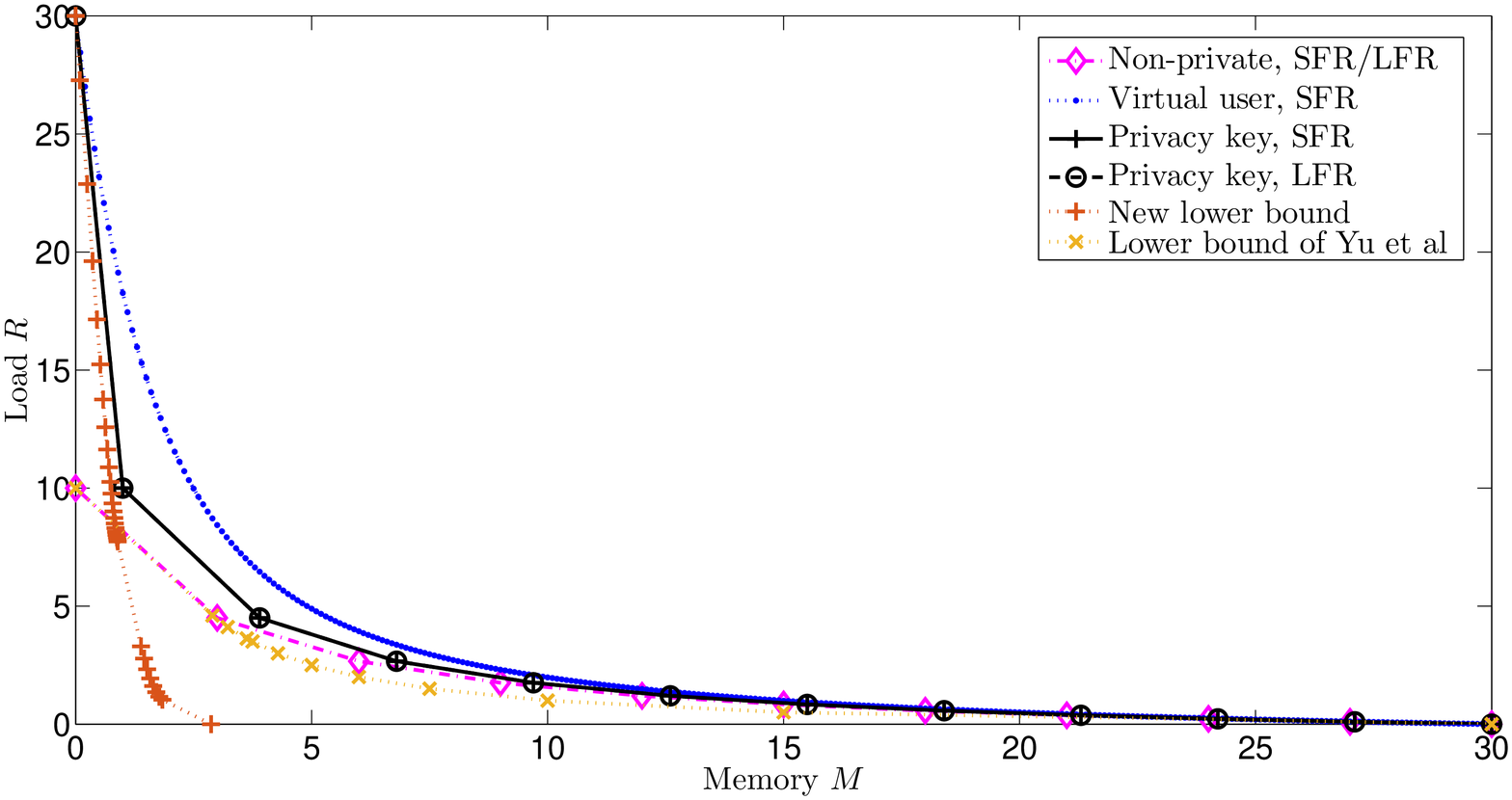}
}
\subfigure[Subpacketization v.s. memory size] {
\label{fig2:b}
\includegraphics[width=0.95\columnwidth]{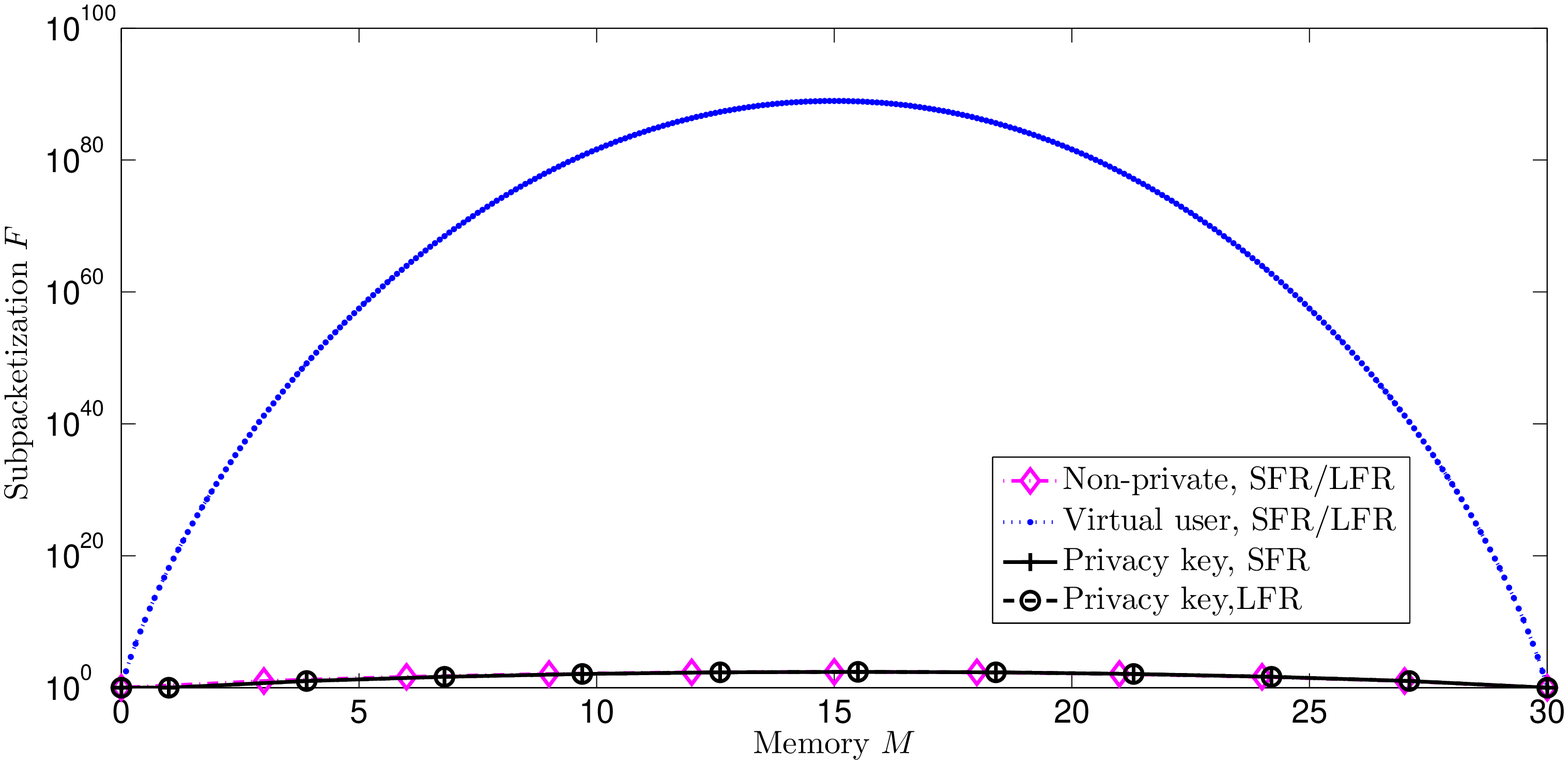}
}

\caption{Performance comparison for an $(N,K)=(30,10)$ system.   }
\label{fig:2}
\end{figure}

The comparisons for  $N\leq K$ and $N>K$ are presented  in Fig.~\ref{fig:1} and Fig.~\ref{fig:2}, respectively.
By observing Fig.~\ref{fig:1} and Fig. ~\ref{fig:2}, we note:
\begin{enumerate}

  \item
  case $N\leq K$ (Fig.~\ref{fig:1}): When $M$ is smaller than some threshold,  the privacy key scheme for LFR demands has  slightly larger load than that for  SFR demands, which  achieves worse load-memory tradeoff than the virtual user SFR scheme. The virtual user scheme for SFR demands  can approach almost the same performance as the non-private schemes. However, the  privacy key  schemes could maintain a similar subpacketization order as the non-private scheme, while the virtual users scheme has a significantly larger subpacketization. 

  \item
  case $N>K$ (Fig.~\ref{fig:2}): the privacy key for LFR demands have same performances as that for SFR scheme, which outperforms the virtual users SFR scheme in both load-memory tradeoff and subpacketization.

\end{enumerate}

We explain intuitively the results in Fig.~\ref{fig1:a} and~\ref{fig2:a} as follows. Firstly,  it has been observed in Section~\ref{sec:LFR:CPA} that the privacy key only has  a slightly larger load than privacy key SFR scheme when $N\leq K$ and $t\leq K-N$. This is caused by the different range of the query vectors $\sd{q}_1,\ldots,\sd{q}_K$ as explained in Section~\ref{subsec:LFR}. Secondly,
both the privacy key and  virtual user schemes are based on the MAN uncoded placement scheme. In  the privacy key  schemes, in addition each user caches some random linear combinations of uncached MAN subfiles. This negative effect on load-memory tradeoff becomes less significant when the number of files $N$ becomes large.
In the virtual user scheme, the server creates multicast signals to satisfy $NK$ users, which includes $K$ real users and $N(K-1)$ virtual users, thus some multicast signals are only useful for virtual users, which increases the load compared to the non-private SFR scheme.
This negative effect on load-memory tradeoff becomes more significant when $N$ becomes large.

The regime where privacy key scheme outperforms the virtual user scheme  for SFR demands can be found by the following observations:
\begin{enumerate}
   \item Both schemes achieve the point $(0,N)$, and their slopes  at $M=0$ are  $-\max\big\{N-K,\frac{2N+1-K}{N+K-1}\big\}$  and $-\frac{N+1}{2}$ respectively. When $N=2K+1$, the two slopes are equal, so $2K+1$ is the threshold of $N$ such that the  privacy key scheme for SFR demands outperforms the virtual user scheme when $M$ is close to $0$.
   \item It was proved in~\cite{Kamath2019} that for $M\geq N-\frac{1}{K}$, the virtual user  scheme achieves the cut-set bound $1-\frac{M}{N}$, while privacy key scheme for SFR demands achieves $1-\frac{M-1}{N-1}$ when $M\geq N-\frac{N-1}{K}$. Thus, the virtual users scheme eventually outperforms the privacy key scheme for SFR demands when $M$ increases to $N$. The load of virtual user scheme is given by $\frac{K(N-M)}{KM+1}$ for $M\in\big\{N-\frac{i}{K}:i\in[0:N-1]\big\}$. Therefore, if $N\geq K+2$ and $M=N-1-\frac{1}{K}$, the loads of the two schemes are equal. So, $M=N-1-\frac{1}{K}$ is the threshold that the virtual user scheme outperforms the privacy key scheme for SFR demands.
\end{enumerate}
These observations, together with extensive numerical results, indicate that when $N>2K+1$ and $0<M<N-1-\frac{1}{K}$, the  privacy key scheme outperforms the virtual user scheme for SFR demands. Notice that for the regime $N-1-\frac{1}{K}\leq M\leq N$, the multiplicative gap of privacy key  scheme (i.e., $R_{\rm{F}}(M)=1-\frac{M-1}{N-1}$) compared to the cut-set bound (i.e., $R\geq1-\frac{M}{N}$) is $\frac{N}{N-1}$, which is very close to one for large $N$.

It is worthy pointing out that, the privacy key scheme is to within a constant multiplicative gap of the optimal load-memory tradeoff 
in all regimes, even in the regime $N\leq K$. From Fig.~\ref{fig1:b} and~\ref{fig2:b}, the privacy key schemes have much lower subpacketization in all parameter regimes, since they are designed to satisfy $K$ users, instead of the $NK$ users as in the virtual user schemes.

From Fig.~\ref{fig1:a} and~\ref{fig2:a}, the bound derived in Theorem~\ref{thm:lowerbound} outperforms the existing bound on an interval beginning with $M=0$ in the case $N>K$. From Theorem~\ref{thm:onesinglegap} (but see also Theorems~\ref{thm:gap} and~\ref{thm:gap:2} in Appendix \ref{sec:app:D} and \ref{App:sec:F}), we know that the lower bound enable us to bound the performance of the privacy key scheme to with a constant multiplicative gap over the interval $[0,1]$.

\section{Conclusions}\label{sec:conclusion}
In this paper, we investigated the coded caching shared-link problem where the demands of users must be protected against any subset of colluding users. 
A privacy key scheme is proposed for Linear Function Retrieval (LFR), and a slightly tighter version for Single File Retrieval (SFR). 
The privacy key scheme is order optimal in all parameter regimes and outperforms existing virtual user schemes for SFR demands in some parameter regime and has much lower subpacketization in general.

\appendix
\subsection{Independence Between the Transmit Signal and the Demands}\label{sec:app:IDX}
\yq{
We prove that the privacy condition~\eqref{eqn:privacy} implies~$I(\sd{d}_{[K]};X\,|\,W_{[N]})=0$.  
The following `conditioned' version of Han's inequalities will be useful.
\begin{lemma} [Han's inequalities~\cite{Inform_Cover_book}]\label{lemma:han} Let $X_0,X_1,X_2,\ldots,X_n$ be $n+1$ random variables,  define
\begin{IEEEeqnarray}{c}
h_k^{(n)} := \frac{1}{{n\choose k}}\sum_{\mathcal{S}\subseteq[n]:|\mathcal{S}|=k}\frac{H(X_\mathcal{S}\,|\,X_0)}{k},  \quad
\\
g_k^{(n)} := \frac{1}{{n\choose k}}\sum_{\mathcal{S}\subseteq[n]:|\mathcal{S}|=k}\frac{H(X_\mathcal{S}\,|\,X_{[n]\backslash\mathcal{S}},X_0)}{k}.
\end{IEEEeqnarray}
Then, $h_1^{(n)}\geq \ldots\geq h_n^{(n)}$ and $g_1^{(n)}\leq \ldots\leq g_{n}^{(n)}$.
\end{lemma}


For any $k\in[K-1]$,
\begin{IEEEeqnarray}{rCl}
\frac{I(\sd{d}_{[K]};X\,|\,W_{[N]})}{K}
&=&\frac{H(D_{[K]}\,|\,W_{[N]})}{K}-\frac{H(D_{[K]}\,|\,X,W_{[N]})}{K}\\
&\stackrel{(a)}{\leq}&\frac{1}{{K\choose k}}\sum_{\mathcal{S}\subseteq[K],|\mathcal{S}|=k}\frac{H(D_{\mathcal{S}}\,|\,W_{[N]})}{k}
-
\frac{H(D_{\mathcal{S}}\,|\,D_{[K]\backslash\mathcal{S}},X,W_{[N]})}{k}\\
&=&\frac{1}{{K\choose k}}\sum_{\mathcal{S}\subseteq[K],|\mathcal{S}|=k}\frac{I(D_{\mathcal{S}};D_{[K]\backslash\mathcal{S}},X\,|\,W_{[N]})}{k}\\
&\leq&\frac{1}{{K\choose k}}\sum_{\mathcal{S}\subseteq[K],|\mathcal{S}|=k}\frac{I(D_{\mathcal{S}};D_{[K]\backslash\mathcal{S}},X,Z_{[K]\backslash\mathcal{S}}\,|\,W_{[N]})}{k}\IEEEeqnarraynumspace\\
&\stackrel{(b)}{=}&0,
\end{IEEEeqnarray}
where $(a)$ follows from Lemma~\ref{lemma:han}, and $(b)$ follows from the privacy condition~\eqref{eqn:privacy}.
}

\subsection{Proof of Lemma~\ref{lemma:conlusion:bound}}\label{sec:appB}
We first prove that the conclusion holds for the case $b=\min\{\ell,K\}$ by induction on $\ell$.
For $\ell=b=1$, the inequalities in~\eqref{eq:exam:leftside}--\eqref{eqn:RM} in Example in Section~\ref{exam:converse}  work for any $\mathcal{B}_1=\{n\}$ and $\mathcal{A}_1=\{k\}$ where $ n\in[N],k\in[K]$.
Thus, the conclusion holds for $\ell=1$.

Now, assume that the conclusion holds for $\ell$ where $\ell\in[N-1]$. Consider the case $\ell+1$, let $b'=\min\{\ell+1,K\}$. Let $\mathcal{B}_{\ell+1}$ and $\mathcal{A}_{b'}$ be any subset of $[N]$ and $[K]$ with cardinalities $\ell+1$ and $b'$ respectively. Let $\{d_j: j\in\mathcal{A}_{b'}\}$ be the demands of users in $\mathcal{A}_{b'}$, which can be any distinct demands in $\mathcal{B}_{\ell+1}$.
We have
\begin{enumerate}
  \item if $\ell<K$, then  $b'=b+1=\ell+1$,  pick any $k\in\mathcal{A}_{b+1}$.
  \begin{IEEEeqnarray}{rCl}
 &&(R+(\ell+1)\cdot  M)B\\
 &=&(R+\ell\cdot M)B+MB\\
 &\stackrel{(a)}{\geq}&H(W_{\mathcal{B}_{\ell+1}\backslash\{d_k\}},X,Z_{\mathcal{A}_{b+1}\backslash\{k\}}\,|\,\{D_j=d_j\}_{j\in\mathcal{A}_{b+1}\backslash\{k\}})+H(Z_{k})\label{proof:induction}\\
 &\stackrel{(b)}{=}&H(W_{\mathcal{B}_{\ell+1}\backslash\{d_k\}},X,Z_{\mathcal{A}_{b+1}\backslash\{k\}}\,|\,\{D_j=d_j\}_{j\in\mathcal{A}_{b+1}})+H(Z_{k}\,|\,\{D_j=d_j\}_{j\in\mathcal{A}_{b+1}}\})\quad\\
  &\geq&H(W_{\mathcal{B}_{\ell+1}\backslash\{d_k\}},X,Z_{\mathcal{A}_{b+1}}\,|\,\{D_j=d_j\}_{j\in\mathcal{A}_{b+1}}\})\\
   &\stackrel{(c)}{=}&H(W_{\mathcal{B}_{\ell+1}},X,Z_{\mathcal{A}_{b+1}}\,|\,\{D_j=d_j\}_{j\in\mathcal{A}_{b+1}}),
 \end{IEEEeqnarray}
where~$(a)$ follows from the induction assumption; $(b)$ follows from~\eqref{eqn:privacyII} and~\eqref{eqn:Zk},~\eqref{eqn:indep.cond.}; and $(c)$ follows from the correctness condition~\eqref{eqn:correctness}.
 
  \item if $\ell\geq K$, then  $b'=b=K<\ell+1$ and  $\mathcal{A}_{b'}=\mathcal{A}_b=[K]$.  Pick any $n\in\mathcal{B}_{\ell+1}\backslash \{d_j:j\in[K]\}$ and $k\in[K]$, let $\{d_j':j\in[K]\}$ be another realization of the demands of users such that
      \begin{IEEEeqnarray}{c}
      d_j'=\Bigg\{\begin{array}{ll}
                    d_j ,&\textnormal{if}~j\neq k  \\
                    n ,&\textnormal{if}~j=k
                  \end{array}
     .\label{d:prime}
      \end{IEEEeqnarray}
      Then,
  \begin{IEEEeqnarray}{rCl}
 &&(R+(\ell+1)\cdot  M)B\\
 &=&(R+\ell\cdot M)B+MB\\
 &\stackrel{(a)}{\geq}&H(W_{\mathcal{B}_{\ell+1}\backslash\{d_k\}},X,Z_{[K]}\,|\,D_{[K]}=d_{[K]}')+H(Z_{k})\label{proof:induction:2}\\
 &\geq&H(W_{\mathcal{B}_{\ell+1}\backslash\{d_k\}},X,Z_{[K]\backslash\{k\}}\,|\,D_{[K]}=d_{[K]}')+H(Z_{k})\\
  &\stackrel{(b)}{=}&H(W_{\mathcal{B}_{\ell+1}\backslash\{d_k\}},X,Z_{[K]\backslash\{k\}}\,|\,D_{[K]}=d_{[K]})+H(Z_{k}\,|\,D_{[K]}=d_{[K]})\\
    &\geq&H(W_{\mathcal{B}_{\ell+1}\backslash\{d_k\}},X,Z_{[K]}\,|\,D_{[K]}=d_{[K]})\\
     &\stackrel{(c)}{=}&H(W_{\mathcal{B}_{\ell+1}},X,Z_{[K]}\,|\,D_{[K]}=d_{[K]}),
 \end{IEEEeqnarray}
  where~$(a)$ follows from the induction assumption; $(b)$ follows from~\eqref{eqn:privacyII} and~\eqref{eqn:Zk},~\eqref{eqn:indep.cond.};  and $(c)$ follows from the correctness condition~\eqref{eqn:correctness}.
\end{enumerate}
Therefore, the conclusion holds for $b=\min\{\ell,K\}$.

For the case $b<\min\{\ell,K\}$, let $\widetilde{\mathcal{A}}\subseteq[K]$ be a subset of size $\min\{\ell, K\}$ such that $\mathcal{A}_b\subset\widetilde{\mathcal{A}}$ and $\{d_j:j\in\widetilde{\mathcal{A}}\backslash\mathcal{A}_b\}$ be any distinct demands of users in $\widetilde{\mathcal{A}}\backslash\mathcal{A}_b$ such that $\{d_j:j\in\widetilde{\mathcal{A}}\backslash\mathcal{A}_b\}\subseteq \mathcal{B}_{\ell}\backslash \{d_j:j\in\mathcal{A}_b\}$. Then $\{d_j:j\in\widetilde{\mathcal{A}}\}$ are distinct demands of users in $\widetilde{\mathcal{A}}$.
 Then by the conclusion for the case $b=\min\{\ell,K\}$,  for sufficiently large $B$,
\begin{IEEEeqnarray}{rCl}(R+\ell\cdot  M)B
&\geq& H(W_{\mathcal{B}_{\ell}},X,Z_{\widetilde{\mathcal{A}}}\,|\,\{D_j=d_j\}_{j\in\widetilde{\mathcal{A}}})\\
&\geq&H(W_{\mathcal{B}_{\ell}},X,Z_{\mathcal{A}_b}\,|\,\{D_j=d_j\}_{j\in\widetilde{\mathcal{A}}})\\
&\stackrel{(a)}{=}&H(W_{\mathcal{B}_{\ell}},X,Z_{\mathcal{A}_b}\,|\,\{D_j=d_j\}_{j\in\mathcal{A}_b}),
\end{IEEEeqnarray}
where $(a)$ follows from the privacy condition~\eqref{eqn:privacyII}.

\subsection{Proof of Lemma~\ref{lemma:h}}\label{sec:appC}
By the definition of $h_\ell(a)$ in~\eqref{eqn:hla}, we have
\begin{IEEEeqnarray}{rCl}
&&h_{\ell}(a)+H(W_{[\ell]},X,Z_{[b_\ell]}\,|\,D_{[b_\ell]}=[b_\ell])\\
&=&\sum_{j=0}^{b_\ell-1}H(W_{[a]},X,Z_{[j]}\,|\,D_{[j]}=[j])+H(W_{[a]},Z_{[b_\ell+1]})+H(W_{[\ell]},X,Z_{[b_\ell]}\,|\,D_{[b_\ell]}=[b_\ell])\IEEEeqnarraynumspace\\
&\stackrel{(a)}{=}&\sum_{j=0}^{b_\ell-1}H(W_{[a]},X,Z_{[j]}\,|\,D_{[j]}=[j])\notag\\
&&\quad +H(W_{[a]},Z_{[b_\ell+1]}\,|\,D_{[b_\ell]}=[b_\ell])+H(W_{[\ell]},X,Z_{[b_\ell]}\,|\,D_{[b_\ell]}=[b_\ell],D_{b_\ell+1}=a+1)\notag\\
&\stackrel{(b)}{\geq}&\sum_{j=0}^{b_\ell-1}H(W_{[a]},X,Z_{[j]}\,|\,D_{[j]}=[j]) +H(W_{[\ell]},Z_{[b_\ell]}\,|\,D_{[b_\ell]}=[b_\ell],D_{b_\ell+1}=a+1)\notag\\
&&\quad +H(W_{[a]},X,Z_{[b_\ell+1]}\,|\,D_{[b_\ell]}=[b_\ell],D_{b_\ell+1}=a+1)\notag\\
&\stackrel{(c)}{=}&\sum_{j=0}^{b_\ell-1}H(W_{[a]},X,Z_{[j]}\,|\,D_{[j]}=[j]) +H(W_{[\ell]},Z_{[b_\ell]}\,|\,D_{[b_\ell]}=[b_\ell],D_{b_\ell+1}=a+1)\notag\\
&&\quad +H(W_{[a+1]},X,Z_{[b_\ell+1]}\,|\,D_{[b_\ell]}=[b_\ell],D_{b_\ell+1}=a+1)\IEEEeqnarraynumspace\\
&\geq&\sum_{j=0}^{b_\ell-1}H(W_{[a]},X,Z_{[j]}\,|\,D_{[j]}=[j])+H(W_{[\ell]},Z_{[b_\ell]}\,|\,D_{[b_\ell]}=[b_\ell],D_{b_\ell+1}=a+1)\notag\\
&&\quad+H(W_{[a+1]},Z_{[b_\ell+1]}\,|\,D_{[b_\ell]}=[b_\ell],D_{b_\ell+1}=a+1)\\
&\stackrel{(d)}{=}&\sum_{j=0}^{b_\ell-1}H(W_{[a]},X,Z_{[j]}\,|\,D_{[j]}=[j])+H(W_{[\ell]},Z_{[b_\ell]}) +H(W_{[a+1]},Z_{[b_\ell+1]}),\label{continue1}
\end{IEEEeqnarray}
where $(a)$ follows from~\eqref{eqn:Zk},~\eqref{eqn:indep.cond.} and~\eqref{eqn:privacyII}; $(b)$ follows from the submodularity property~\eqref{eqn:submodualrity}; $(c)$ follows from the correction condition~\eqref{eqn:correctness}; and $(d)$ follows from~\eqref{eqn:Zk} and~\eqref{eqn:indep.cond.}.

Notice that, for any $j\in[0:b_\ell-1]$, we have
\begin{IEEEeqnarray}{rCl}
&&H(W_{[a]},X,Z_{[j]}\,|\,D_{[j]}=j)+H(W_{[\ell]},Z_{[j+1]})\\
&\stackrel{(a)}{=}&H(W_{[a]},X,Z_{[j]}\,|\,D_{[j]}=j,D_{j+1}=a+1)+H(W_{[\ell]},Z_{[j+1]}\,|\,D_{[j]}=j,D_{j+1}=a+1)\\
&\stackrel{(b)}{\geq}&H(W_{[a]},X,Z_{[j+1]}\,|\,D_{[j]}=j,D_{j+1}=a+1)+H(W_{[\ell]},Z_{[j]}\,|\,D_{[j]}=j,D_{j+1}=a+1)\\
&\stackrel{(c)}{=}&H(W_{[a+1]},X,Z_{[j+1]}\,|\,D_{[j]}=j,D_{j+1}=a+1)+H(W_{[\ell]},Z_{[j]}\,|\,D_{[j]}=j,D_{j+1}=a+1)\IEEEeqnarraynumspace\\
&\geq&H(W_{[a+1]},X,Z_{[j]}\,|\,D_{[j]}=j,D_{j+1}=a+1)+H(W_{[\ell]},Z_{[j]}\,|\,D_{[j]}=j,D_{j+1}=a+1)\\
&\stackrel{(d)}{=}&H(W_{[a+1]},X,Z_{[j]}\,|\,D_{[j]}=j)+H(W_{[\ell]},Z_{[j]}),\label{continue2}
\end{IEEEeqnarray}
where $(a)$ follows from~\eqref{eqn:Zk},~\eqref{eqn:indep.cond.} and~\eqref{eqn:privacyII};  $(b)$ follows from the submodularity property~\eqref{eqn:submodualrity};  $(c)$ follows from the correction condition~\eqref{eqn:correctness}; and $(d)$ follows from~\eqref{eqn:privacyII} and~\eqref{eqn:Zk},~\eqref{eqn:indep.cond.}.

Thus, we can continue with~\eqref{continue1},
\begin{IEEEeqnarray}{rCl}
&&h_{\ell}(a)+H(W_{[\ell]},X,Z_{[b_\ell]}\,|\,D_{[b_\ell]}=[b_\ell])\\
&\stackrel{(a)}{\geq}&\sum_{j=0}^{b_\ell-2}H(W_{[a]},X,Z_{[j]}\,|\,D_{[j]}=[j])+H(W_{[\ell]},Z_{[b_\ell-1]})\notag\\
&&\quad+H(W_{[a+1]},X,Z_{[b_\ell-1]}\,|\,D_{[b_\ell-1]}=[b_\ell-1]) +H(W_{[a+1]},Z_{[b_\ell+1]})\\
&&\quad\vdots\label{eq:omit}\\
&\stackrel{(b)}{\geq}&H(W_{[\ell]})+\sum_{j=0}^{b_\ell-1}H(W_{[a+1]},X,Z_{[b_\ell-1]}\,|\,D_{[b_\ell-1]}=[b_\ell-1]) +H(W_{[a+1]},Z_{[b_\ell+1]})\\
&\stackrel{(c)}{=}&h_\ell{(a+1)}+H(W_{[\ell]}),
\end{IEEEeqnarray}
where from $(a)$ to $(b)$, we  apply~\eqref{continue2} to $j=b_\ell-2,b-3,\ldots,0$ sequentially; and $(c)$ follows from the definition of $h_\ell(a)$ in~\eqref{eqn:hla}.

\subsection{Gap Results for SFR Demands}\label{sec:app:D}

\begin{theorem}\label{thm:gap}
For an $(N,K)$ DPCU system, the ratio of the achieved communication loads of the privacy key scheme for SFR demands  $R_{\rm{F}}(M)$ and the optimal communication load for SFR demands $R_{\rm{F}}^*(M)$ is upper bounded by
\begin{IEEEeqnarray}{c}
\frac{R_{\rm{F}}(M)}{R_{\rm{F}}^*(M)}\leq\left\{\begin{array}{ll}
                                  2 ,&\textnormal{if}~ 0\leq M\leq 1,\, N\geq 2K \\
                                  2, &\textnormal{if}~ 0\leq M\leq \frac{1}{2},\, N<2K\\
                                  4, &\textnormal{if}~\frac{1}{2}\leq M\leq 1,\, N< 2K\\
                                  4, &\textnormal{if}~1\leq M\leq N, N\geq\frac{K(K+1)}{2}\\
                                 4.0177,  &\textnormal{if}~1\leq M\leq N, K<N<\frac{K(K+1)}{2}\\
                                 5.4606, &\textnormal{if}~1\leq M\leq N, N\leq K
                                \end{array}
\right..\label{eqn:R:gap}
\end{IEEEeqnarray}
\end{theorem}

\begin{IEEEproof}
We bound $\frac{R_{\rm{F}}(M)}{R_{\rm{F}}^*(M)}$ for  $0\leq M\leq 1$ and $1\leq M\leq N$ separately.
\subsubsection{Case $0\leq M\leq 1$}\label{app:subM} For clarity, we denote the function in the braces of~\eqref{eqn:lower:bound} by $f(M,\ell)$, i.e.,
\begin{IEEEeqnarray}{c}
f(M,\ell) := \ell+\frac{\min\{\ell+1,K\}\cdot(N-\ell)}{N-\ell+\min\{\ell+1,K\}}-\ell\cdot  M,\quad\forall\, \ell\in[N], M\in[0,N].
\end{IEEEeqnarray}

We further discuss in two subcases $N\geq 2K$ and $N<2K$.
\begin{enumerate}
   \item If $N\geq 2K$, $R_{\rm{F}}(M)$ is upper bounded by the line segment connecting $(0,N)$ and $(M_0,R_0)=(1,K)$, i.e.,
  \begin{IEEEeqnarray}{c}
  R_{\rm{F}}(M)\leq L(M) :=  N-(N-K)M, \quad M\in[0,1].\label{eqn:RA:bound}
  \end{IEEEeqnarray}
  We further discuss in two sub-cases, i.e.,  $0\leq M\leq 1-\frac{K}{N}$ and $ 1-\frac{K}{N}\leq M\leq 1$.
  \begin{enumerate}
    \item If $0\leq M\leq 1-\frac{K}{N}$,
      \begin{IEEEeqnarray}{rCl}
  \frac{R_{\rm{F}}(M)}{R_{\rm{F}}^*(M)}\leq\frac{L(M)}{f(M,N)}
  \stackrel{(a)}{\leq} \frac{L(1-\frac{K}{N})}{f(1-\frac{K}{N},N)}
  =2-\frac{K}{N}
  \leq2,
  \end{IEEEeqnarray}
   where in $(a)$ we utilized the fact that $\frac{L(M)}{f(M,N)}$ increases with $M$ over $\big[0,1-\frac{K}{N}\big]$.
    \item If $1-\frac{K}{N}\leq M\leq 1$,
    \begin{IEEEeqnarray}{rCl}
    \frac{R_{\rm{F}}(M)}{R_{\rm{F}}^*(M)}\leq\frac{L(M)}{f(M,K)} 
\stackrel{(a)}{\leq}\max_{M\in\{1-\frac{K}{N},1\}}\Big\{\frac{L(M)}{f(M,K)}\Big\}
=\max\Big\{2-\frac{K}{N}, \frac{1}{1-\frac{K}{N}}\Big\}
\stackrel{(b)}{\leq}2.\IEEEeqnarraynumspace
\end{IEEEeqnarray}
where~$(a)$ holds because for any fixed $(N,K)$ such that $N\geq 2K$, the linear fractional function $\frac{L(M)}{f(M,K)}$ is either increasing or decreasing in $M$ over $\big[1-\frac{K}{N},1\big]$, and~$(b)$ follows from the fact $\frac{K}{N}\leq\frac{1}{2}$.
  \end{enumerate}

  \item If $N<2K$, $R_{\rm{F}}(M)$ is upper bounded by $N$.
We further discuss in two sub-cases, i.e.,  $0\leq M\leq \frac{1}{2}$ and $ \frac{1}{2}<M\leq 1$.
 \begin{enumerate}
 \item If $0\leq M\leq\frac{1}{2}$,
 \begin{IEEEeqnarray}{c}
 \frac{R_{\rm{F}}(M)}{R_{\rm{F}}^*(M)}\leq \frac{N}{f(M,N)}=\frac{N}{N-NM}
 \leq 2.
 \end{IEEEeqnarray}
 \item If $\frac{1}{2}<M\leq 1$,
 \begin{IEEEeqnarray}{rCl}
 \frac{R_{\rm{F}}(M)}{R_{\rm{F}}^*(M)}\stackrel{(a)}{\leq}\frac{N}{f(1,\lfloor N/2\rfloor)}
 =\frac{N+1}{\lfloor N/2\rfloor+1}\cdot\frac{N}{N-\lfloor N/2\rfloor}
 \stackrel{(b)}{\leq} 4,\label{exp:last}
 \end{IEEEeqnarray}
 where in~$(a)$ follows since $f(M,\lfloor N/2\rfloor)$ decreases with $M$, and in $(b)$, we used the fact $\frac{N}{2}\geq\lfloor N/2\rfloor\geq \frac{N}{2}-\frac{1}{2}$.
 \end{enumerate}

\end{enumerate}

\subsubsection{Case $1\leq M\leq N$}
Denote the optimal centralized and decentralized load for an $(N,K)$ system with  memory size $M$ at each user under uncoded placement without privacy constraint by $r_{\rm C}(M)$ and $r_{\rm D}(M)$ respectively. By the results of~\cite{Yu2019ExactTradeoff}, $r_{\rm C}(M)$ is the lower convex envelope of the  points $(M,R)$  in the first column of~Table~\ref{table:1}
      and $r_{\rm D}(M)$ is given by
      \begin{IEEEeqnarray}{c}
      r_{\rm D}(M) :=  \frac{N-M}{M}\Big(1-\Big(1-\frac{M}{N}\Big)^{\min\{N,K\}}\Big),\quad \forall\, M\in[0,N].
      \end{IEEEeqnarray}
      Denote the optimal load for an $(N,K)$ system with  memory size $M$ at each user without privacy constraint by $r^{*}(M)$. By the results in~\cite{QYu2018Factor2},
      \begin{IEEEeqnarray}{c}
       r_{\rm C}(M)\leq r_{\rm D}(M)\leq\left\{\begin{array}{ll}
                                             2.00884 \cdot r^*(M), &\textnormal{if}~N<\frac{K(K+1)}{2} \\
                                             2\cdot r^*(M), & \textnormal{if}~N\geq \frac{K(K+1)}{2}
                                           \end{array}
        \right.. \label{eqn:ineq:chain}
      \end{IEEEeqnarray}
      Notice that the optimal load for SFR is no less than that without privacy constraint, i.e.,
      \begin{IEEEeqnarray}{c}
      r^*(M)\leq R_{\rm{F}}^*(M),\quad\forall\, M\in[0,N].\label{eqn:opt:comp}
      \end{IEEEeqnarray}

   We bound $\frac{R_{\rm{F}}(M)}{R_{\rm{F}}^*(M)}$ for the subcases $N>K$ and $N\leq K$ separately.
      \begin{enumerate}
        \item If $N>K$, we bound $\frac{R_{\rm{F}}(M)}{R_{\rm{F}}^*(M)}$ for the sub-cases $(N,K)=(3,2)$ and $(N,K)\neq (3,2)$ separately.
        \begin{enumerate}
        \item If $(N,K)=(3,2)$, the lower convex envelope of the points $(0,N)=(0,3)$, $(M_0,R_0)=(1,2)$, $(M_1,R_1)=(2,\frac{1}{2})$ and $(M_2,R_2)=(3,0)$ is given by
            \begin{IEEEeqnarray}{c}
            R_{\rm{F}}(M)=\Bigg\{\begin{array}{ll}
                            3-\frac{5}{4}M ,&\textnormal{if}~0\leq M\leq 2  \\
                            \frac{3}{2}-\frac{1}{2}M ,&\textnormal{if}~2\leq M\leq 3
                          \end{array}
            .
            \end{IEEEeqnarray}
            By Theorem~\ref{thm:lowerbound} and the cut set bound $NR+M\geq N$ (see~\cite{Maddah-Ali2014fundamental}), we have $R_{\rm{F}}^*(M)\geq 1-\frac{1}{3}M$, thus
            \begin{enumerate}
            \item If $1\leq M\leq 2$,
            \begin{IEEEeqnarray}{rCl}
            \frac{R_{\rm{F}}(M)}{R_{\rm{F}}^*(M)}\leq\frac{3-\frac{5}{4}M}{1-\frac{1}{3}M}\leq \frac{21}{8}<4.
            \end{IEEEeqnarray}
            \item For $2\leq M< 3$,
          \begin{IEEEeqnarray}{rCl}
          \frac{R_{\rm{F}}(M)}{R_{\rm{F}}^*(M)}&\leq&\frac{\frac{3}{2}-\frac{1}{2}M}{1-\frac{1}{3}M}=\frac{3}{2}<4.
          \end{IEEEeqnarray}
            \end{enumerate}

        \item If $(N,K)\neq (3,2)$,  we prove the following inequality in Appendix~\ref{sub:app:C},
        \begin{IEEEeqnarray}{c}
        \frac{R_{\rm{F}}(M)}{r_{\rm C}(M)}\leq 2,\quad 1\leq M\leq N.\label{eqn:RC:ratio}
        \end{IEEEeqnarray}
 Therefore,
        \begin{IEEEeqnarray}{rCl}
        \frac{R_{\rm{F}}(M)}{R_{\rm{F}}^*(M)}&=&\frac{R_{\rm{F}}(M)}{r_{\rm C}(M)}\cdot\frac{r_{\rm C}(M)}{r^*(M)}\cdot\frac{r^*(M)}{R_{\rm{F}}^*(M)}
        \leq2\cdot\frac{r_{\rm C}(M)}{r^*(M)}\\
        &\stackrel{(a)}{<}&\Bigg\{\begin{array}{ll}
                    4,    &\textnormal{if}~N\geq\frac{K(K+1)}{2}  \\
                    4.0177,    &\textnormal{if}~K< N<\frac{K(K+1)}{2}
                     \end{array}
        ,
        \end{IEEEeqnarray}
where $(a)$ follows from~\eqref{eqn:ineq:chain}.
        \end{enumerate}

        \item If $N\leq K$,   we prove the following inequality in Appendix~\ref{sub:app:C},
        \begin{IEEEeqnarray}{c}
        \frac{R_{\rm{F}}(M)}{r_{\rm D}(M)}\leq e,\quad 1\leq  M\leq N. \label{eqn:RD:ratio}
        \end{IEEEeqnarray}

        Therefore,
        \begin{IEEEeqnarray}{rCl}
        \frac{R_{\rm{F}}(M)}{R_{\rm{F}}^*(M)}
        &=&\frac{R_{\rm{F}}(M)}{r_{\rm D}(M)}\cdot\frac{r_{\rm D}(M)}{r^*(M)}\cdot\frac{r^*(M)}{R_{\rm{F}}^*(M)}\IEEEeqnarraynumspace\\
        &\leq& e\times 2.00884\times 1
        <5.4606.\label{eqn:ratio:RF}
        \end{IEEEeqnarray}
      \end{enumerate}
\end{IEEEproof}

\subsection{Gap Results for LFR Demands}\label{App:sec:F}
\yq{

\begin{theorem}\label{thm:gap:2}
For an $(N,K)$ DPCU system, the ratio of the achieved communication loads of the privacy key  scheme for LFR demands $R_{\rm{L}}(M)$ and the optimal communication load for LFR demands $R_{\rm{L}}^*(M)$ is upper bounded by
\begin{IEEEeqnarray}{c}
\frac{R_{\rm{L}}(M)}{R_{\rm{L}}^*(M)}\leq\left\{\begin{array}{ll}
                                  2 ,&\textnormal{if}~ 0\leq M\leq 1,\, N\geq 2K \\
                                  2, &\textnormal{if}~ 0\leq M\leq \frac{1}{2},\, N<2K\\
                                  4, &\textnormal{if}~\frac{1}{2}\leq M\leq 1,\, N< 2K\\
                                  4, &\textnormal{if}~1\leq M\leq N, N\geq\frac{K(K+1)}{2}\\
                                 4.0177,  &\textnormal{if}~1\leq M\leq N, K<N<\frac{K(K+1)}{2}\\
                                 6.3707, &\textnormal{if}~1\leq M\leq N, N\leq K
                                \end{array}
\right..\label{eqn:R:gap}
\end{IEEEeqnarray}
\end{theorem}

\begin{IEEEproof}
We prove the theorem in three cases.
\subsubsection{Case $N>K$} By~\eqref{eqn:achievable} and~\eqref{eqn:achievable:LFR}, in this case $R_t=R_t'$ for all $t\in[0:K]$. Therefore, $R_{\rm{L}}(M)=R_{\rm{F}}(M)$ for  $M\in[0,N]$. On the other hand, $R_{\rm{L}}^*(M)\geq R_{\rm{F}}^*(M)$ for $M\in[0,N]$, then the conclusion directly follows from the conclusion in Theorem~\ref{thm:1}.

\subsubsection{Case $N\leq K$, $0\leq M\leq 1$} The proof directly follows from the same lines as in the proof for the subcase $0\leq M\leq 1, N<2K$ in Appendix~\ref{sec:app:D}, except that $R_{\rm{F}}(M)$ and $R_{\rm{F}}^*(M)$ are replaced by $R_{\rm{L}}(M)$ and $R_{\rm{L}}^*(M)$ respectively.
\subsubsection{Case $N\leq K$, $1\leq M\leq N$} Notice that the lower convex envelope of $\{(M_t,R_t):t\in[0:K]\}$ is  formed by connecting $(M_0,R_0),(M_1,R_1),\ldots,(M_K,R_K)$ sequentially. Thus, $R_{\rm{F}}(M)$ is formed by connecting the points $(0,N), (M_{t_{\rm{F}}},R_{t_{\rm{F}}}),(M_{t_{\rm{F}}+1},R_{t_{\rm{F}}+1}),\ldots,(M_K,R_K)$ sequentially for some $t_{\rm{F}}\in[0:K]$. Similarly, $R_{\rm{L}}(M)$ is formed by connecting the points $(0,N),(M_{t_{\rm{L}}},R_{t_{\rm{L}}}'),(M_{t_{\rm{L}}+1},R_{t_{\rm{L}}+1}'),\ldots,(M_K,R_K')$ for some $t_{\rm{L}}\in[0:K]$. As a result, the maximum value of $\frac{R_{\rm L}(M)}{R_{\rm F}(M)}$ must be obtained at some point in $\{M_t:t\in[0:K]\}$. Notice that $R_t=R_t'$ for $t\geq K-N+1$, therefore
\begin{IEEEeqnarray}{rCl}
\frac{R_{\rm L}(M)}{R_{\rm F}(M)}&\leq &\max_{t\in[0:K-N]}\Big\{\frac{R_t'}{R_t}\Big\}\\
&=&\max_{t\in[0:K-N]}\bigg\{\frac{{K\choose t+1}-{K-N\choose t+1}}{{K\choose t+1}-{K-N+1\choose t+1}}\bigg\}\\
&=&\max_{t\in[0:K-N]}\bigg\{\frac{\sum_{s=1}^N{K-s\choose t}}{\sum_{s=1}^{N-1}{K-s\choose t}}\bigg\}\\
&=&\max_{t\in[0:K-N]}\bigg\{1+\frac{{K-N\choose t}}{\sum_{s=1}^{N-1}{K-s \choose t}}\bigg\}\\
&\leq &1+\frac{1}{N-1}.\label{eqn:ratio:R0}
\end{IEEEeqnarray}
Now consider two cases.
\begin{enumerate}
\item If $N\leq 6$, $R_{\rm{L}}(M)$ is upper bounded by $N-M$, and $R_{\rm L}^*(M)$ is lower bounded by $1-\frac{M}{N}$ by cut set bound~\cite{Maddah-Ali2014fundamental}, thus
\begin{IEEEeqnarray}{c}
\frac{R_{\rm L}(M)}{R_{\rm L}^*(M)}\leq \frac{N-M}{1-M/N}=N\leq 6<6.3707.
\end{IEEEeqnarray}
\item If $N\geq 7$,
\begin{IEEEeqnarray}{rCl}
\frac{R_{\rm L}(M)}{R_{\rm L}^*(M)}&=&\frac{R_{\rm L}(M)}{R_{\rm F}(M)}\cdot \frac{R_{\rm F}(M)}{R_{\rm F}^*(M)}\cdot \frac{R_{\rm F}^*(M)}{R_{\rm L}^*(M)}\\
&\stackrel{(a)}{\leq} &\Big(1+\frac{1}{N-1}\Big)\times e\times 2.00884\times 1\\
&\stackrel{(b)}{\leq} &6.3707,
\end{IEEEeqnarray}
where $(a)$ follows from~\eqref{eqn:ratio:R0},~\eqref{eqn:ratio:RF} and $(b)$ follows from $N\geq 7$.
\end{enumerate}
\end{IEEEproof}
}

\subsection{Proof of the Two Inequalities in \eqref{eqn:RC:ratio} and \eqref{eqn:RD:ratio}}\label{sub:app:C}

For notational clarity, in this subsection, we will use $r_{\rm C}(M,N,K)$ and $r_{\rm D}(M,N,K)$ to denote the function  $r_{\rm C}(M)$ and $r_{\rm D}(M)$  in \eqref{eqn:RC:ratio} and \eqref{eqn:RD:ratio} respectively, which are defined for an $(N,K)$ system. 
We aim to prove the inequalities~\eqref{eqn:RC:ratio} and~\eqref{eqn:RD:ratio}, i.e.,
\begin{lemma}\label{lem:ineq}
\begin{enumerate} For an $(N,K)$ system,
  \item If $N>K$ and $(N,K)\neq (2,3)$,
\begin{IEEEeqnarray}{c}
        \frac{R_{\rm{F}}(M)}{r_{\rm C}(M,N,K)}\leq 2,\quad 1\leq M\leq N.\label{ineq:1}
\end{IEEEeqnarray}
  \item If $N\leq K$,
  \begin{IEEEeqnarray}{c}
   \frac{R_{\rm{F}}(M)}{r_{\rm D}(M,N,K)}\leq e,\quad 1\leq M\leq N. \label{ineq:2}
   \end{IEEEeqnarray}
\end{enumerate}
\end{lemma}

\begin{IEEEproof}
By Theorem~\ref{thm:1}, for $M\geq 1$, $R_{\rm{F}}(M)$ is uppper bounded by the lower convex envelope of the points $\{(M_t,R_t):t\in[0:K]\}$ in~\eqref{eqn:achievable}, which is exactly $r_{\rm C}(M-1,N-1,K)$ by Table~\ref{table:1}. Thus, by~\eqref{eqn:ineq:chain}, for $1\leq M\leq N$,
      \begin{IEEEeqnarray}{c}
      R_{\rm{F}}(M)\leq r_{\rm C}(M-1,N-1,K)\leq r_{\rm D}(M-1,N-1,K).\label{eqn:chain:2}
      \end{IEEEeqnarray}
Therefore,
\begin{enumerate}
\item If $N>K$ and $(N,K)\neq(3,2)$, since $R_{\rm{F}}(M)$ is convex in $M$, and $r_{\rm C}(M,N,K)$ is  piecewise linear in $M$ with corner points such that $M\in\{1\}\cup\{\frac{tN}{K}:t\in[K]\}$ over $[1,N]$, it suffices to prove~\eqref{ineq:1} for the cases $M\in\{1\}\cup\{\frac{tN}{K}:t=1,\ldots,K\}$.
        \begin{enumerate}
        \item If $M=1$, let $\theta=1-\frac{K}{N}$, then $M=1=\theta\cdot 0+(1-\theta)\frac{N}{K}$, thus
            \begin{IEEEeqnarray}{rCl}
            r_{\rm C}(1,N,K)&=&\theta\cdot r_{\rm C}(0,N,K)+(1-\theta)\cdot r_{\rm C}\Big(\frac{N}{K},N,K\Big)\\
            &=&K\Big(1-\frac{K+1}{2N}\Big)\\
            &\geq&\frac{K}{2},\label{eqn:rc}
            \end{IEEEeqnarray}
           where we used the fact $N\geq K+1$. Thus,
            \begin{IEEEeqnarray}{c}
            \frac{R_{\rm{F}}(1)}{r_{\rm C}(1,N,K)}\stackrel{(a)}{\leq}\frac{r_{\rm C}(0,N-1,K)}{r_{\rm C}(1,N,K)}\stackrel{(b)}{\leq} 2,
            \end{IEEEeqnarray}
where $(a)$ follows from~\eqref{eqn:chain:2}, and $(b)$ follows from $r_{\rm C}(0,N-1,K)=K$ and~\eqref{eqn:rc}.
            \item If $M=\frac{N}{K}$ and $N=K+1$, $R_{\rm{F}}(M)$ is upper bounded by the line segment connecting $(0,N)$ and $(M_1,R_1)=(\frac{K+N-1}{K}, \frac{K-1}{2})$, i.e.,
                \begin{IEEEeqnarray}{c}
                R_{\rm{F}}(M)\leq N-\frac{K(2N+1-K)}{2(K+N-1)}M,\quad\forall\, M\in\Big[0,\frac{K+N-1}{K}\Big].\IEEEeqnarraynumspace\label{eqn:L3}
                \end{IEEEeqnarray}
                 Thus,
                \begin{IEEEeqnarray}{rCl}
                \frac{R_{\rm{F}}(\frac{N}{K})}{r_{\rm C}(\frac{N}{K},N,K)}&\leq&\frac{N-\frac{K(2N+1-K)}{2(K+N-1)}\cdot\frac{N}{K}}{\frac{K-1}{2}}=\frac{3(K+1)}{2K}\leq2,
                \end{IEEEeqnarray}
                where we used the fact $K\geq 3$ since $(N,K)\neq(3,2)$.

             \item If $M=\frac{tN}{K}$, where  $t\geq 2$ or $N\geq K+2$ hold, we have
             \begin{IEEEeqnarray}{c}
             M-1=\theta\Big(\frac{(t-1)(N-1)}{K}\Big)+(1-\theta)\Big(\frac{t(N-1)}{K}\Big),\label{eqn:comb}
             \end{IEEEeqnarray}
              where $\theta=\frac{K-t}{N-1}<1$. Thus,
              \begin{IEEEeqnarray}{rCl}
              &&\frac{R_{\rm{F}}(M)}{r_{\rm C}(M,N,K)}\\
              &\stackrel{(a)}{\leq}&\frac{r_{\rm C}(M-1,N-1,K)}{r_{\rm C}(M,N,K)}\\
              &\stackrel{(b)}{=}&\frac{\theta \cdot r_{\rm C}\big(\frac{(t-1)(N-1)}{K},N-1,K\big)+(1-\theta)\cdot r_{\rm C}\big(\frac{t(N-1)}{K},N-1,K\big)}{r_{\rm C}(M,N,K)}\\
              &=&\frac{\frac{K-t}{N-1}\cdot\frac{K-t+1}{t}+\frac{N-1-K+t}{N-1}\cdot\frac{K-t}{t+1}}{\frac{K-t}{t+1}}\\
              &=&1+\frac{K+1}{(N-1)t}\\
              &\stackrel{(c)}{\leq}&2,
              \end{IEEEeqnarray}
              where $(a)$ follows from~\eqref{eqn:chain:2}; $(b)$ follows from~\eqref{eqn:comb}; and in $(c)$, we used the fact $t\geq 2$ or $N\geq K+2$.
    \end{enumerate}
\item If $N\leq K$, let $q :=  1-\frac{M}{N}\in\big[0,1-\frac{1}{N}\big]$, then $1-\frac{M-1}{N-1}=\frac{N}{N-1}q$. Hence,
        \begin{IEEEeqnarray}{rCl}
         \frac{R_{\rm{F}}(M)}{r_{\rm D}(M,N,K)}&\stackrel{(a)}{\leq}&\frac{r_{\rm D}(M-1,N-1,K)}{r_{\rm D}(M,N,K)}\\
         &=&\frac{N}{N-1}\cdot\frac{\frac{M}{N}}{\frac{M-1}{N-1}}\cdot\frac{1-\big(1-\frac{M-1}{N-1}\big)^{N-1}}{1-\big(1-\frac{M}{N}\big)^N}\\
         &=&\frac{N}{N-1}\cdot\frac{1-q}{1-\frac{N}{N-1}q}\cdot\frac{1-(\frac{N}{N-1})^{N-1}q^{N-1}}{1-q^N}\\
         &=&\frac{N}{N-1}\cdot\frac{1+\frac{N}{N-1}q+\ldots+\big(\frac{N}{N-1}\big)^{N-2}q^{N-2}}{1+q+\ldots+q^{N-1}}\IEEEeqnarraynumspace\\
         &\leq&\Big(\frac{N}{N-1}\Big)^{N-1}\cdot\frac{1+q+\ldots+q^{N-2}}{1+q+\ldots+q^{N-1}}\\
         &\leq&\Big(1+\frac{1}{N-1}\Big)^{N-1}\\
         &<&e,
        \end{IEEEeqnarray}
where in $(a)$, we used~\eqref{eqn:chain:2}.

        \end{enumerate}
 \end{IEEEproof}

\end{document}